\documentclass[aps,pra,nobalancelastpage,superscriptaddress, 10pt, twocolumn]{revtex4-2}

\usepackage{color,ifthen,amsthm,amsmath,amsxtra,amsfonts,dsfont,graphicx,bm,tikz,scalerel,wasysym,bbm,graphicx,amsthm,physics,empheq,comment, soul}
\usepackage[colorlinks=true,linkcolor=blue, citecolor=blue, urlcolor=blue, bookmarks]{hyperref}

\newcommand{\be}{\begin{equation}}
\newcommand{\ee}{\end{equation}}
\newcommand{\bea}{\begin{eqnarray}}
\newcommand{\eea}{\end{eqnarray}}
\newcommand{\bse}{\begin{subequations}}
\newcommand{\ese}{\end{subequations}}
\newcommand{\Np}{N'}
\newcommand{\Ap}{A'}

\theoremstyle{plain}
\newcommand{\1}{\mathbbm{1}}

\theoremstyle{plain}

\theoremstyle{plain}

\usepackage{color}
\definecolor{myred}{RGB}{232,102,102}
\definecolor{myblue}{RGB}{187,187,255}
\definecolor{myorange0}{RGB}{252,226,5}
\definecolor{myorange0c}{RGB}{255,255,255}
\definecolor{myorange}{RGB}{255,165,0}
\definecolor{mygrey}{RGB}{105,105,105}
\definecolor{OliveGreen}{RGB}{85,107,47}
\definecolor{NavyBlue}{RGB}{0,0,128}
\definecolor{mygreen}{RGB}{34,139,34}
\definecolor{myY}{RGB}{220,255,203}
\definecolor{myYO}{RGB}{255, 220, 151}
\definecolor{mygreenc}{RGB}{150,50,50}

\newcommand{\Xe}[2]{
\draw[thick] (#1-0.5/2, #2) -- (#1+0.5/2,#2);
\draw[thick] (#1-0.5/2,#2-.5/2) -- (#1-0.5/2,#2+0.5/2);
\draw[thick] (#1+0.5/2,#2-.5/2) -- (#1+0.5/2,#2+0.5/2);
\draw[ fill=myblue, rounded corners=2.5] (#1+0.5/2-0.125,#2-0.125) rectangle (#1+0.5/2+0.125,#2+0.125);
}

\newcommand{\XeBig}[2]{
\draw[] (#1-0.5*2, #2) -- (#1+0.5*2,#2);
\draw[] (#1-0.5*2,#2-.45) -- (#1-0.5*2,#2+0.5);
\draw[] (#1+0.5*2,#2-.45) -- (#1+0.5*2,#2+0.5);
\draw[ fill=myblue, rounded corners=2.5] (#1+1-0.25/2,#2-0.25/2) rectangle (#1+1+0.25/2,#2+0.25/2);
}

\newcommand{\XeOne}[2]{
\draw[thick] (#1+0.5/2,#2-.5/2) -- (#1+0.5/2,#2+0.5/2);
\draw[ fill=myblue, rounded corners=2.5] (#1+0.5/2-0.125,#2-0.125) rectangle (#1+0.5/2+0.125,#2+0.125);
}

\newcommand{\btp}{\begin{tikzpicture}[baseline=(current  bounding  box.center), scale=.7] }
\newcommand{\btps}{\begin{tikzpicture}[baseline=(current  bounding  box.center), scale=.4] }
\newcommand{\etp}{\end{tikzpicture}}

\usetikzlibrary{shapes.geometric}

\hypersetup{
pdftitle={Localised Dynamics in the Floquet Quantum East Model},
pdfsubject={Many-body quantum systems},
pdfauthor={Bruno Bertini, Pavel Kos, and Toma\v z Prosen},
pdfkeywords={quantum circuits, quantum many-body systems, localisation}
}

\usepackage{xr} 
\newcommand{\titleinfo}{Localised Dynamics in the Floquet Quantum East Model}

\begin{document}
\title{\titleinfo}

\author{Bruno Bertini}
\affiliation{School of Physics and Astronomy, University of Nottingham, Nottingham, NG7 2RD, UK}
\affiliation{Centre for the Mathematics and Theoretical Physics of Quantum Non-Equilibrium Systems,
University of Nottingham, Nottingham, NG7 2RD, UK}

 \author{Pavel Kos}
 \affiliation{Max-Planck-Institut f\"ur Quantenoptik, Hans-Kopfermann-Str. 1, 85748 Garching}
 
\author{Toma\v z Prosen}
 \affiliation{Department of Physics, Faculty of Mathematics and Physics, University of Ljubljana, Jadranska 19, SI-1000 Ljubljana, Slovenia}

\begin{abstract}

We introduce and study the discrete-time version of the Quantum East model, an interacting quantum spin chain inspired by simple kinetically constrained models of classical glasses. Previous work has established that its continuous-time counterpart displays a disorder-free localisation transition signalled by the appearance of an exponentially large (in the volume) family of non-thermal, localised eigenstates. Here we combine analytical and numerical approaches to show that: i) The transition persists for discrete times, in fact, it is present for any finite value of the time step apart from a zero measure set; ii) It is directly detected by following the non-equilibrium dynamics of the fully polarised state. Our findings imply that the transition is currently observable in state-of-the-art platforms for digital quantum simulation. 
\end{abstract}

\maketitle

\textit{Introduction.---} Establishing the precise conditions for real space localisation in interacting systems, even in one dimension, turns out to be extremely challenging. Despite intense efforts to crack it~\cite{altshuler,huse,dimalioms,ros,avalanches,reviewbloch,vidmar,abanini2021distinguishing,sels}, it currently remains a major unsolved problem in theoretical physics. It has been argued that a form of many-body localisation should emerge as a consequence of an external quenched disorder, which, under some conditions, might defeat interactions. Whether this mechanism can lead to a stable phase of matter remains an actively debated topic~\cite{vidmar, abanini2021distinguishing, sels}. A fundamental problem is that localisation studies are either limited to small systems accessible to numerical or experimental simulation or uncontrolled perturbative approximations. Nevertheless, for many-body localisation to be established as a phase of matter it has to exist in the thermodynamic limit: it should not (only) be a property of eigenstates, but (also) of dynamics. 

Recently, it has been suggested that, other than by disorder, real space localisation can also be triggered by kinetic constraints which render transport a higher-order process. 
An advantage of this approach is its immunity to fluctuating rare events such as ergodic bubbles.
A minimal example of this mechanism is realised in the so-called Quantum East model~\cite{horssen2015dynamics, crowley2017entanglement, roy2020strong, brighi2022hilbert,geissler2022slow,klobas2023exact} (and its bosonic version \cite{valencia2022kinetically}) where a localisation transition in the quantum Hamiltonian is in one-to-one correspondence to a first order activity-inactivity transition in the corresponding classical stochastic glass model. In agreement with this picture, Ref.~\cite{pancotti2020quantum} observed an eigenstate localisation transition in the Quantum East for an exponentially large family of eigenstates.

In this work, we take a fundamental step further and look for the possibility of dynamical localisation in a Floquet, or trotterised, version of the Quantum East model, where localisation is challenged by a steady pumping of energy into the system~\cite{lazarides2014equilibrium, dalessio2014long, ponte2015periodically}. This setting can be seen as the kinetically-constrained analogue of Floquet many-body localisation~\cite{ponte2015many, lazarides2015fate, abanin2016theory}. We replace the continuous Hamiltonian dynamics by a discrete sequence of conditional unitary gate operations -- a quantum circuit -- that can be conveniently implemented on platforms for digital quantum simulation such as trapped ions~\cite{lanyon2011universal,barreiro2011open,blatt2012quantum, monroe2021programmable} and superconducting circuits~\cite{salathe2015digital,barends2015digital,langford2017experimentally,wendin2017quantum, kjaergaard2020superconducting, bravyi2022future}. 
Using time-dependent perturbation theory, we argue that the model displays a localisation transition by tuning the 
parameters of the model. We demonstrate that in the dynamically localised phase the model can be efficiently simulated by time-dependent matrix product methods (i.e.\ TEBD algorithm)~\cite{vidal,schollwock2011density, cirac2021matrix} to an arbitrary precision, showing very good quantitative agreement with the  perturbative prediction.
Moreover, we find qualitative agreement between the dynamical picture of localisation in the infinite system and the localisation of eigenstates in the finite system.

\textit{The model.---} 
Our starting point is the Quantum East model~\cite{horssen2015dynamics} defined by the following Hamiltonian operator
(in arbitrary energy units)
\be
H(a)= \sum_{j=1}^{2L-1} P_j (a X_{j+1}-I) + a X_{1}-I\,. 
\label{eq:Hquatumeast}
\ee
Here $a$ is the dimensionless coupling constant, $2L$ is the system size, $\{X_j, Y_j, Z_j\}$ are Pauli matrices acting non-trivially at site $j$, $I$ is the identity operator, and $P_j = (I+Z_j)/2$. 

\begin{figure}[t]
\includegraphics[width=0.4\textwidth]{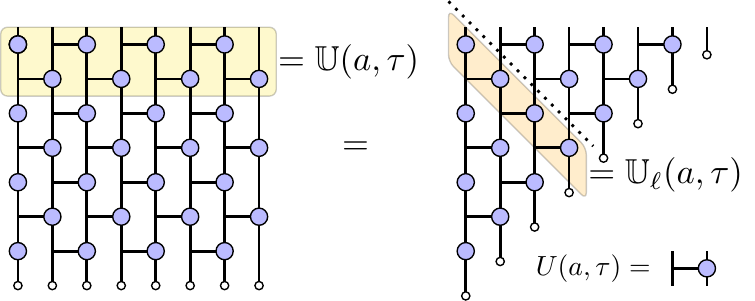}
  \vspace{-.3cm}
    \caption{(Left) State \eqref{eq:initialstate} after $t=3.5$ time steps of discrete dynamics.
    White bullets denote the state $\ket{\downarrow}$ and the blue circles the activation part of the conditional gate $U(a,\tau)$.
    The brick-wall Floquet propagator $\mathbb{U}(a,\tau)$ is highlighted in yellow. 
    (Right) Explicit simplification of the dynamics out of the light cone. Dashed lines indicate the cut for the ladder evolution in Eq.~\eqref{eq:Wladder}. The ladder propagator $\mathbb{U}_{\ell}(a,\tau)$ (cf.~\eqref{eq:ladder}) is highlighted in orange. 
    }
    \label{fig:circuit}
\end{figure}

We are interested in discrete sequences of unitary operations $\mathbb U(a,\tau)$ that reproduce the dynamics generated by Eq.~\eqref{eq:Hquatumeast} in a special scaling limit. Namely
\be
\lim_{t\to\infty}\mathbb U (a,\frak t/t)^t = e^{- i H(a) \frak t}\,,
\label{eq:trotterlimit}
\ee
where $t$ is the number of discrete time steps, while $\frak t$ plays the role of physical time. This procedure is known as Trotter-Suzuki decomposition~\cite{suzuki1991general, trotter1959product} and does not uniquely specify the unitary operator: there are many choices of $\mathbb U(a,\tau)$ fulfilling Eq.~\eqref{eq:trotterlimit}. Here we consider one that is local in space, i.e., it has the following brickwork structure (see Fig.~\ref{fig:circuit})
\be
\mathbb U (a,\tau) = e^{i \tau} \mathbb U_{\rm e}(a,\tau)\mathbb U_{\rm o}(a,\tau),
\label{eq:Floquet}
\ee
with 
\begin{align}
&\mathbb U_{\rm e}(a,\tau)=U_{1,2}(a,\tau)\cdots U_{2L-1,2L}(a,\tau),\notag\\
&\mathbb U_{\rm o}(a,\tau)=  e^{- i \tau a  X_1} U_{2,3}(a,\tau) \cdots U_{2L-2,2L-1}(a,\tau),
\end{align}
where we use a standard notation $O_{x,y}$ for an operator $O$ acting non-trivially only at the sites $x, y$, and define a local conditional gate
$U(a,\tau)=e^{- i \tau(a P\otimes X-P\otimes I)}$.
The (Trotter) time step $\tau$, usually referred to as the Trotter step, sets the strength of the unitary operation \eqref{eq:Floquet}. It is easy to verify that \eqref{eq:trotterlimit} holds for the evolution operator defined in Eq.~\eqref{eq:Floquet}. Note that the discrete-time dynamics generated by Eq.~\eqref{eq:Floquet} is equivalent to a continuous-time dynamics in the presence of a periodic drive.

To probe the localisation properties of the quantum circuit~\eqref{eq:Floquet} we consider a \emph{local quantum quench}. Namely, we prepare the circuit in the initial state 
\be
\ket{\downarrow \cdots \downarrow},
\label{eq:initialstate}
\ee
which is an eigenstate of the bulk evolution due to the identity  
$
{U(a,\tau)\ket{\downarrow\downarrow}=\ket{\downarrow\downarrow}},
$
but is not stationary at the left boundary. As a consequence, only the sites within a light cone spreading from the left boundary undergo a non-trivial evolution, see Fig.~\ref{fig:circuit}. Intuitively, one can think of our local quench protocol as creating a localised disturbance in $(x,t)=(0,0)$ in a state that is otherwise stationary. Importantly, this quench problem is also a caricature of local-operator spreading in a generic quantum many-body system after an operator-to-state mapping. 
Here, $\ket{\downarrow}$ [and (\ref{eq:initialstate})] represents the identity and $\ket{\uparrow}$ stands for any other traceless local operator that starts growing from the left edge. The question of localization now translates to that of the existence of a  conserved local operator.

A simple measure of how the disturbance created by the local quench spreads through the system is given by the partial norms 
\be
W(x,t)\!=\!\!\! \sum_{s_j=\uparrow, \downarrow} |\!\mel{s_1\ldots s_{x-1} \!\!\uparrow \downarrow \!\cdots\!\! \downarrow }{\mathbb U (a,\tau)^t}{\downarrow \!\cdots\! \downarrow}\!|^2\!.
\label{eq:W}
\ee
 Since ${W(x,t)\geq 0}$ and ${\sum_{x} W(x,t) =1}$, the partial norms can be thought of as the probability of having the rightmost up spin at position $x$. Specifically, whenever the disturbance remains localised at the boundary we have $W(x,t)\approx 0$ for $x\gg x_0=O(t^0)$, while when it spreads through the light cone the partial norms attain non-zero values for all $x\leq 2t$. We emphasise that due to the light cone ${W(x>2t,t)=0}$. The factor $2$ is a direct consequence of the brickwork structure of \eqref{eq:Floquet}, as each time step propagates information for up to two sites to the right.     

In fact, to facilitate our numerical analysis we consider slightly modified quantities that bear the same physical information as those in Eq.~\eqref{eq:W}: Instead of the partial norms of the state $\mathbb U (a,\tau)^t \ket{\downarrow \!\cdots\! \downarrow}$, we look at those of the state along the diagonal cut in the right panel of Fig.~\ref{fig:circuit}. The latter quantities are 
\be
N(x,t)\!=\!\!\!\!\! \sum_{s_j=\uparrow, \downarrow} \!\!|\!\mel{s_1\ldots s_{x-1} \!\!\uparrow \downarrow \!\cdots\!\! \downarrow }{\mathbb U_{\ell} (a,\tau)^t}{\downarrow \!\cdots\! \downarrow}\!|^2\!\!,
\label{eq:Wladder}
\ee
where we introduced the ``ladder evolution operator'' (cf. Fig~\ref{fig:circuit})
\be 
\,\,\!\!\!\!\mathbb U_{\ell}(a,\!\tau\!) \!=\!  e^{i \tau}\! e^{-i \tau a  X_1}
U_{1,2}(a,\!\tau\!)
U_{2,3}(a,\!\tau\!)
\!\cdots\! U_{L-1,L}(a,\!\tau\!),
\label{eq:ladder}
\ee 
which is related to $\mathbb U(a,\tau)$ by a similarity transformation~\cite{Note1}. The quantities in Eq.~\eqref{eq:Wladder} are more convenient than those in Eq.~\eqref{eq:W} because with the same computational effort one can access times that are twice as long.

\textit{Infinite system at finite times.---} Let us begin considering the time evolution of the partial norms $N(x,t)$ in the thermodynamic limit $L\to\infty$. In this case, the main qualitative features of their evolution can be understood by performing a simple perturbative analysis (the same can be done for $W(x,t)$~\cite{Note1}). We begin by introducing the interaction representation of the time evolution operator 
 \be
{\mathbb U}_{\ell} (a,\tau)^t = \left[\prod_{k=0}^{t} \tilde {\mathbb U}_{\ell} (\tau a,\tau k) \right] e^{i \tau t \sum_{j} \! P_j }, 
 \label{eq:interactionpicture}
 \ee
 where we defined 
\begin{align} 
& \tilde{\mathbb U}_{\ell}(a,\tau)=  e^{i \tau} e^{- i a  X_1 e^{-i\tau Z_1}} \!\!\!\prod_{k\in\{1,\ldots 2L-1\}}^{\rightarrow}\!\!\! \tilde U_{k,k+1}(a,\tau), 
\end{align}
with $ \tilde U(a, \tau)=e^{- i a P\otimes X e^{- i \tau Z}}$. We now fix $x,t,\tau$ and expand \eqref{eq:W} in powers of $a$. Looking at the local gate in the interaction picture, i.e.,   
\be
\tilde U_{1,2}(a,\tau)=\1 - i a P_1 X_2 e^{- i \tau Z_2} + O(a^2)\,
\label{eq:gateleading}
\ee 
we have that $N(x)$ is at least of order $a^{2x}$. Indeed, due to the structure of Eq.~\eqref{eq:Floquet}, to get a spin up at position $x$ we need to at least flip all the spins on its left. This also tells us 
\begin{align}
\!\!\!\!\!N(x,t) \!\simeq\! \Np(x,t) \!\equiv\!  |\!\mel*{\underbrace{\uparrow\cdots \uparrow}_x \downarrow \!\cdots\! \downarrow }{\mathbb U (a,\tau)^t}{\downarrow \!\cdots\! \downarrow}\!|^2\!\!,
\label{eq:simp}
\end{align}
where $\simeq$ denotes equality at the leading order in $a$. A simple combinatorial calculation then allows us to express it in terms of $q$-deformed binomial coefficients~\cite{Note1}   
\be
N(x,t) \simeq \Np_{1}(x,t) \equiv (a\tau)^{2x} \biggl |\!\begin{pmatrix}
t \\
x
\end{pmatrix}_{\!\!\!q}\biggr |^2\!\!,
\label{eq:WPT1st}
\ee   
where we set $q=e^{i \tau}$. Interestingly, the perturbative analysis commutes with the limit \eqref{eq:trotterlimit}. Indeed 
$
\lim_{\tau\to0}  \Np_{1}(x,\frak t/\tau)  = 
{(2a\sin(\frak t/2))^{2x}}/{(x!)^2},
$
coincides with the leading order of \eqref{eq:Wladder} if one replaces \eqref{eq:interactionpicture} with its Trotter limit~\cite{Note1}.

Let us now move on to analyse the localisation properties of the perturbative solution. To this end, we assume that $\Np_{1}(x,t)$ gives the only relevant contribution to the partial norm. 
The first key feature of $\Np_{1}(x,t)$ is that its localisation properties depend on whether or not $\tau$ is a rational multiple of $2\pi$. Namely, whether it can be written as ${2\pi c}/{d}$ for some coprime integers $c$ and $d$. When true,  the $q$-Lucas theorem~\cite{sagan1992congruence} connects the behaviour of $q$-deformed and regular binomials
\be
\begin{pmatrix}
t \\
x
\end{pmatrix}_{\!\!\! q} = \begin{pmatrix}
\lfloor t/d\rfloor \\
\lfloor x/d\rfloor
\end{pmatrix} \begin{pmatrix}
{\rm mod}(t,d) \\
{\rm mod}(x,d) 
\end{pmatrix}_{\!\!\!q}\,,
\label{eq:qlucas}
\ee
where $\lfloor x \rfloor$ is the largest integer smaller than $x$ and ${\rm mod}(c,d)$ is the remainder of the division of $c\in\mathbb N$ by $d\in\mathbb N$. Using the Stirling approximation we find that Eq.~\eqref{eq:WPT1st} has a maximum at $
\bar x = 
{(a\tau)^{2d} t}/{(1+(a\tau)^{2d})}
$. Therefore the support of $\Np_{1}(x,t)$ grows in time ruling out localisation.    

On the other hand, whenever $\tau$ is not a rational multiple of $2\pi$ the deformed binomial coefficients are bounded in time. Namely we have 
\begin{align}
\log \left |\begin{pmatrix}
t \\
x
\end{pmatrix}_{\!\!\!q}\right |^2  = & \sum_{p=1}^{t-x} \log\left[\frac{1-\cos(\tau(x+p))}{1-\cos(\tau p)}\right] \simeq O(t^0).
\label{eq:qbinomialbound}
\end{align}
In the last step we used that, since $\{{\rm mod}(\tau p, 2\pi)\}_{p=1}^t$ covers $[0,2\pi)$ uniformly in the large $t$ limit, we have 
\begin{align}
\sum_{p=1}^{t-x} \log(1-\cos(\tau(y+p)))\simeq (x-t) \log 2\,,\quad \forall y\,. 
\end{align}
Therefore the $O(t)$ in Eq.~\eqref{eq:qbinomialbound} cancels, and we are left with an $O(t^0)$ term. Plugging the bound Eq.~\eqref{eq:qbinomialbound} into Eq.~\eqref{eq:WPT1st} we find that $\Np_{1}(x,t)$ is localised within a distance $x_0 = {- 1}/{(2\log a\tau)}$ from the left boundary for all times. 

The second key feature is the $\tau$ dependence of $a_c$ --- the critical $a$ for localisation --- in the case of irrational $\tau/(2\pi)$. In our setup this amounts to ask for what range of $a$ we expect the perturbative result to apply (at least qualitatively). From Eq.~\eqref{eq:WPT1st} we see that for finite $\tau$ the parameter that has to be small to ensure the validity of the perturbative approach is $a\tau$. Instead, in the limit $\tau\to0$ the perturbative solution requires $a$ itself to be small~\cite{Note1}. This suggests that $a_c$ should be of the form 
\be
a_c(\tau)=\min(\alpha, \beta/\tau)\,,
\label{eq:actworegimes}
\ee
for some $\alpha,\beta\in \mathbb R$. 

\begin{figure}    
\hbox{\hspace{-0.465cm}\includegraphics[width=0.53\textwidth]{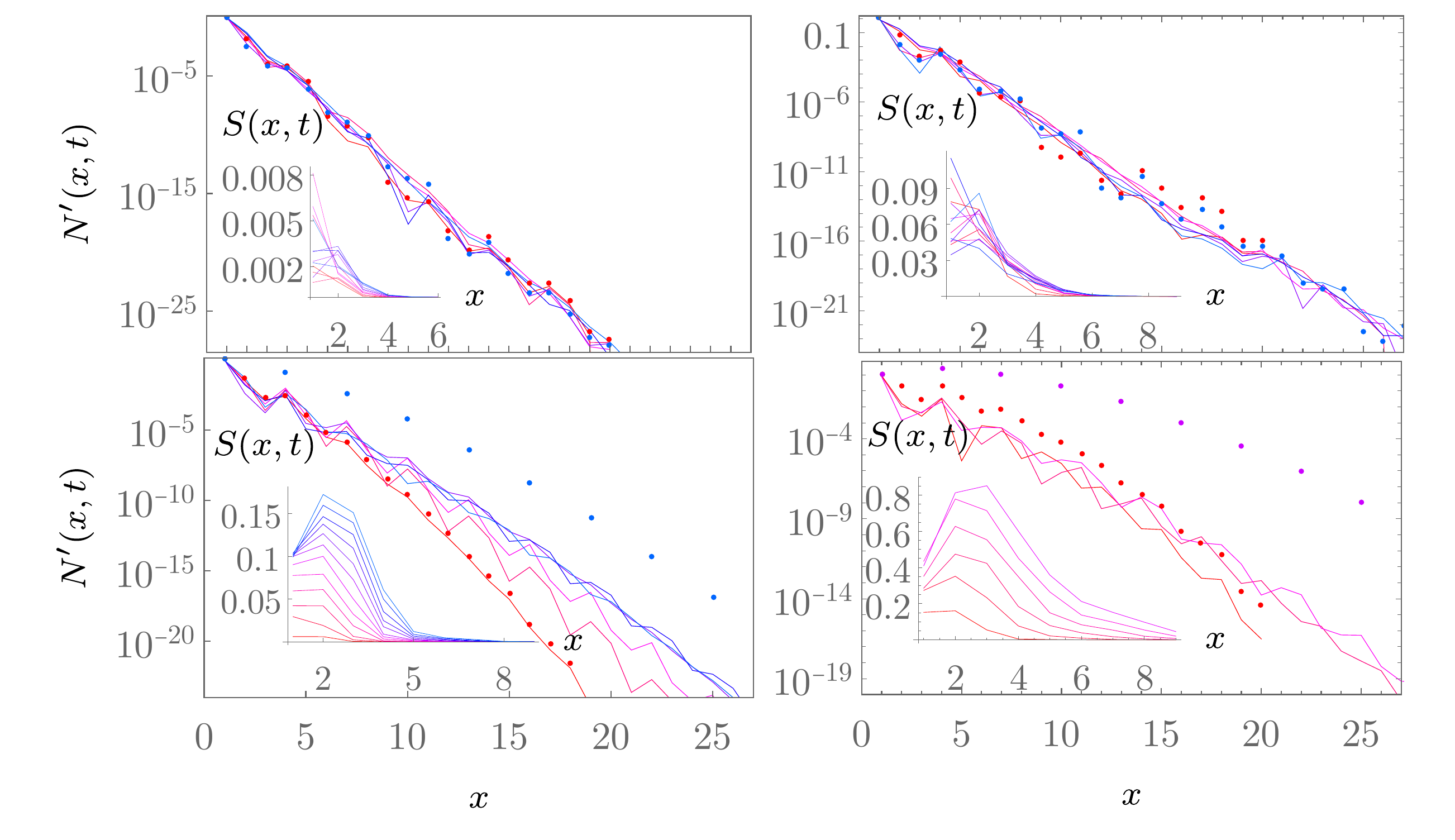}}
  \vspace{-3mm}
    \caption{
Profiles of $\Np(x,t)$ for $a=0.1$ (left column panels), $a=0.2$ (right column panels), and for $\tau=(\sqrt{5}-1)\pi/2$ (top row panels), $\tau = 2\pi/3$ (bottom row panels). The data are shown for times $t=20,40\ldots 120$ (red to blue curves), except for bottom/right panel where only $t=20,40,60$ can be computed due to fast growth of entanglement (insets indicate entanglement entropy profiles $S(x,t)$, $t=10,20\ldots$, of respective cases). Coloured bullets depict perturbative results for shortest and longest simulated time.}
 \label{fig:infTloc}
\end{figure}

\begin{figure}   
  \hbox{\hspace{-0.33cm}\includegraphics[width=0.52\textwidth]{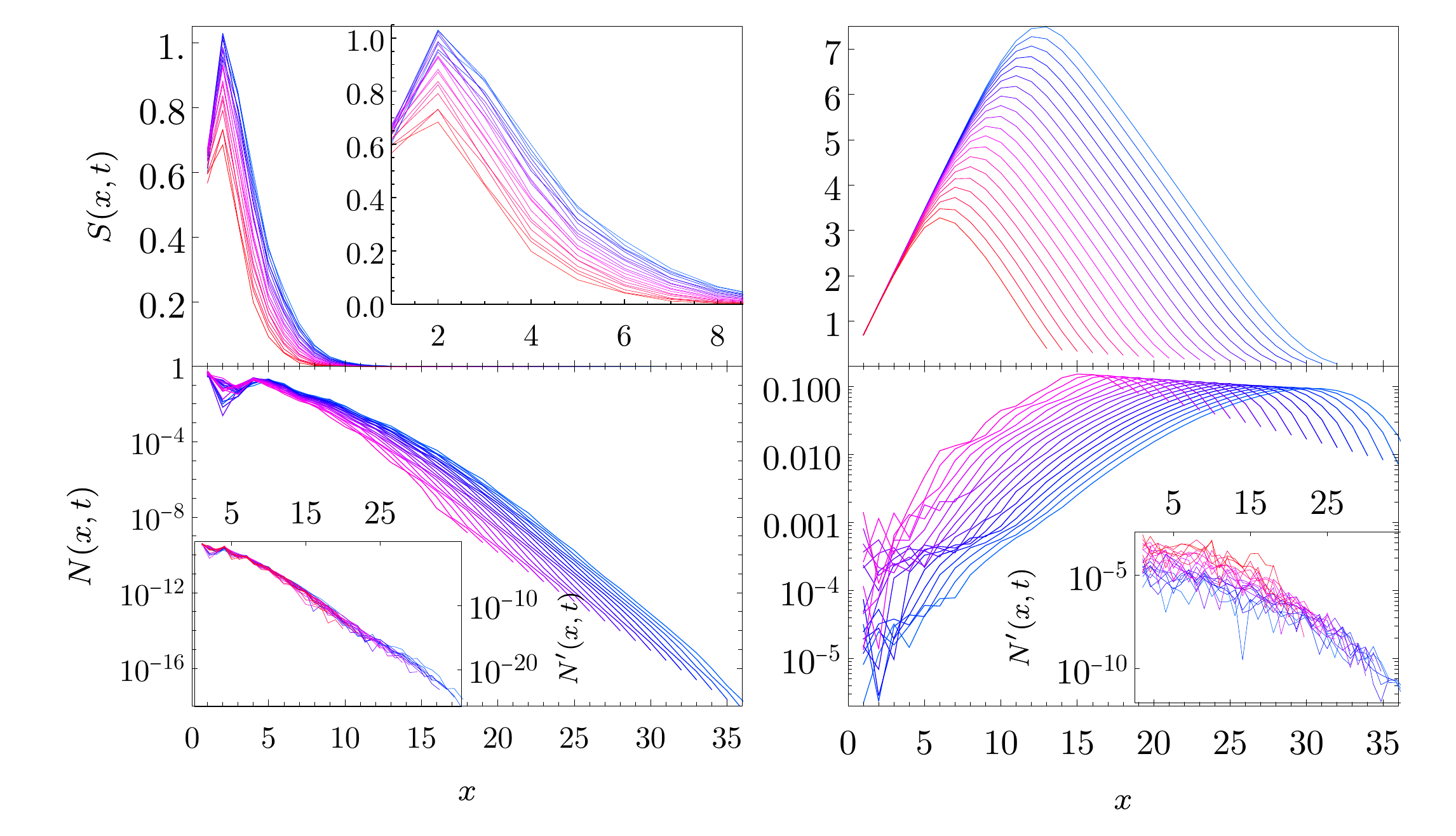}}
  \vspace{-.4cm}
    \caption{
    Two cases of ergodic finite $t$/infinite size dynamics: left, right, column panels correspond, respectively, to $a=0.3,\tau=(\sqrt{5}-1)\pi/2$ just beyond localization  transition, and to $a=1.0,\tau=(\sqrt{5}-1)\pi/2$ well in the ergodic phase. 
    We plot
 entanglement entropy profiles $S(x,t)$, partial norm profiles $N(x,t)$, and domain wall component profiles $\Np(x,t)$ (insets) for $t=17,18\ldots 36$ (red to blue).}
    \label{fig:infTerg}
\end{figure}

Remarkably, by computing $N(x,t)$ and $\Np(x,t)$ via a simple version of the TEBD algorithm~\cite{Note1} we find that all these qualitative features persist away from the perturbative regime. Some representative examples of our numerical results are presented in Figs.~\ref{fig:infTloc} and~\ref{fig:infTerg}, where, as a further indicator of localisation, we also report the entanglement entropy $S(x,t)$ between the $x$ leftmost sites and the rest of the system at time $t$. 

For small enough $a$ we see that disturbance created by the local quench remains localised only for irrational values of $\tau/(2\pi)$. This is clearly shown in the insets of Fig.~\ref{fig:infTloc}: While for rational $\tau/(2\pi)$ we see the peak (and its position) of the entanglement entropy growing linearly in time, for irrational $\tau/(2\pi)$ we see it saturating (additional corroborating plots of the (spatio) temporal behaviour of $S(x,t)$, $W(x,t)$ and entanglement spectra are found in \cite{Note1}). Note that we observe this stark difference between rational and irrational $\tau/(2\pi)$ also for times that are significantly out of the perturbative regime ($t\gg1/a$) and at which Eq.~\eqref{eq:WPT1st} is not quantitatively accurate: see the comparison in the main panel of Fig.~\ref{fig:infTloc}. As a result of this localised behaviour, for irrational $\tau/(2\pi)$ we are able to run our TEBD simulations with essentially no truncation error for hundreds of time steps.  

On the other hand, for $a$ larger than a certain critical value $a_c(\tau)$ the system transitions to the ergodic regime also for irrational $\tau/(2\pi)$, see Fig.~\ref{fig:infTerg}. In this case the perturbative result does not describe the system's behaviour even at the qualitative level: the support of the partial norms grows linearly in time signalling a delocalisation of the disturbance caused by the impurity. Concerning the $\tau$ dependence of $a_c(\tau)$, our numerical results are compatible with the functional form in Eq.~\eqref{eq:actworegimes}~\cite{Note1}. Namely, the critical $a$ appears approximately $\tau$-independent for small $\tau$
, while it starts to decay as $\tau^{-1}$ when $\tau$ is increased beyond a threshold value.

\textit{Finite systems at infinite times.---} 
\begin{figure}[t]
    \centering
    \includegraphics[width=0.42\textwidth]{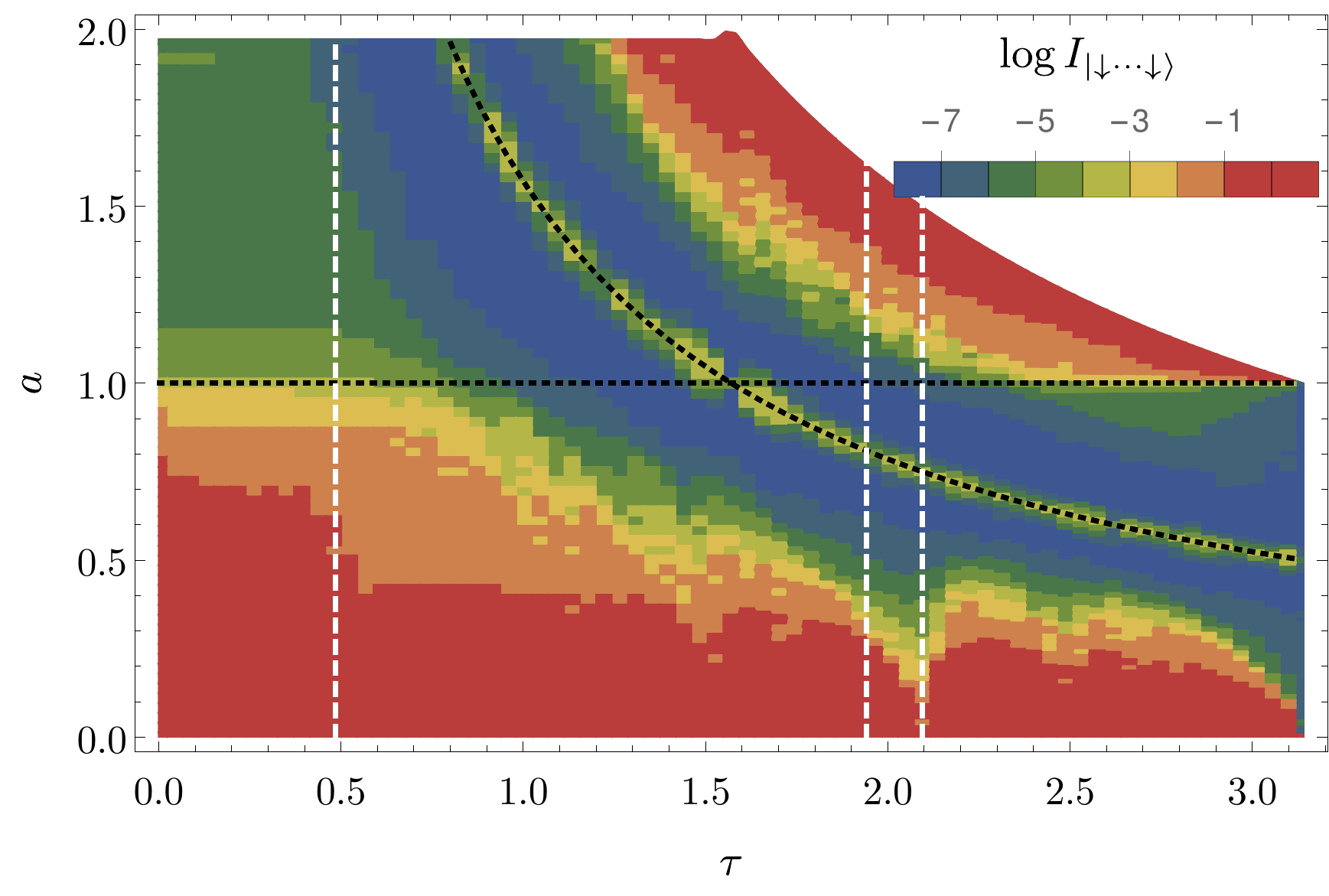} 
    \includegraphics[width=0.42\textwidth]{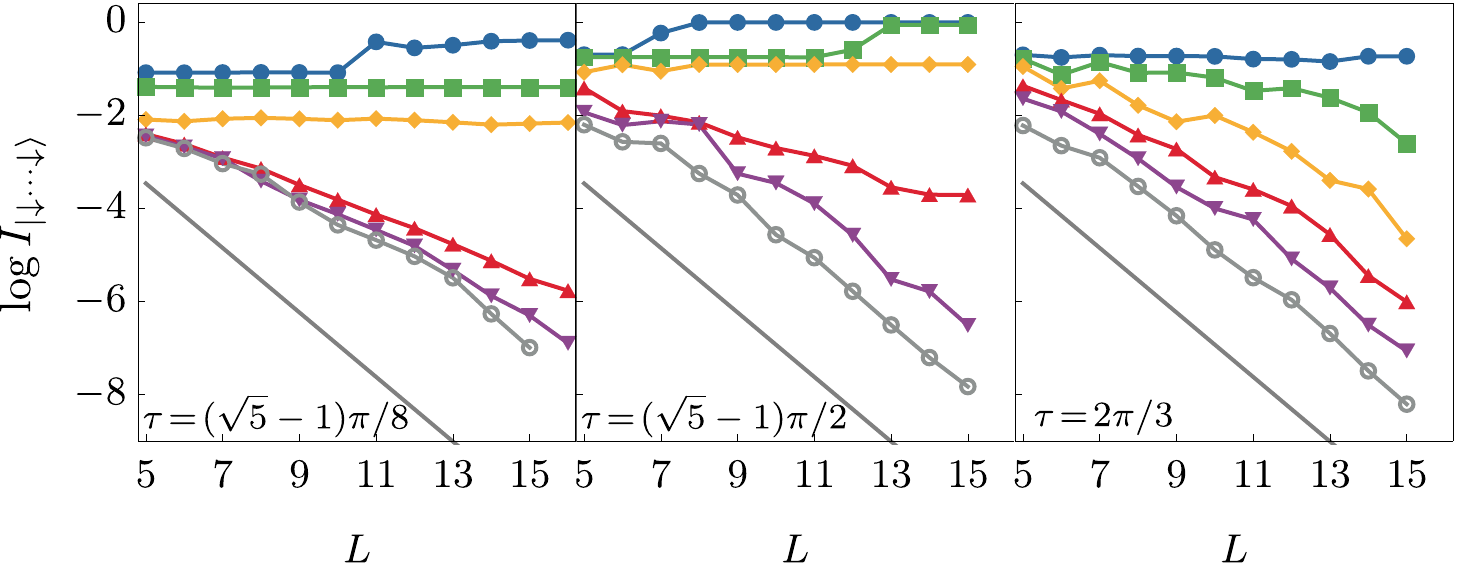}
      \vspace{-.4cm}
    \caption{
    (Top) Logarithm of the IPR at $L=13$ as a function of $\tau$ and $a$. We see the transition for smaller (bigger) $\tau$  at $a=1$ ($a\propto 1/\tau$). The dashed lines denote the three $\tau$s considered in the bottom panel. (Bottom) Logarithm of the IPR versus $L$ for three values of $\tau$ and several values of $a$ ($0.5,0.7,\dots,1.5$ top to bottom for the first plot and $0.05,0.15,\dots,0.55$ for the second and third). The solid grey line corresponds to random eigenstates. For the first two $\tau$s the transition occurs around $a=1$ and $a=0.3$. The third $\tau$ is a rational multiple of $2\pi$, and its transition occurs at a much smaller $a$. For a detailed analysis of this plot, see \cite{Note1}.}
    \label{fig:2dPD}
\end{figure}
Interestingly, the phenomenology observed above in the thermodynamic limit is also observed for finite volumes. Here we again look at a quench from the initial state in Eq.~\eqref{eq:initialstate} but keep $L$ finite while taking ${t \to \infty}$. A convenient indicator of the localisation transition is then the time averaged square of Loschmidt Echo (LE) $|\bra{\downarrow \cdots \downarrow}\mathbb{U}_{\ell}^t \ket{\downarrow \cdots \downarrow}|^2$ [this quantity is the same for brick-wall $\mathbb U$ and ladder propagators $\mathbb{U}_{\ell}$]. Assuming that there are no degeneracies in the spectrum of $\mathbb{U}_{\ell}$, the LE can be written as 
$
\lim_{t \to \infty} \frac{1}{t} \sum_{s}^t  |\bra{\downarrow \cdots \downarrow}
{\mathbb{U}_{\ell} (a,\tau)}^{s}\ket{\downarrow \cdots \downarrow}|^2 
= \sum_{i} |\bra{\downarrow \cdots \downarrow}\ket{E_i}|^4 \equiv I_{\ket{\downarrow \cdots \downarrow}}
$, 
where the sum over $i,j$ goes over all eigenstates. $I_{\ket{\downarrow \cdots \downarrow}}$ is the inverse participation ratio (IPR), which measures the spreading of the initial state in the eigenbasis of the time-evolution operator. It can be interpreted as the purity of the probability distribution $\{P_i\}$, with $P_i=|\bra{\downarrow \cdots \downarrow}\ket{E_i}|^2$ being the Born probability of measuring the eigenstate $\ket{E_i}$ in $\ket{\downarrow \cdots \downarrow}$. 
For random eigenstates $\ket{E_i}$ the probability distribution is flat, i.e., $P_i={2^{-L}}$, which gives $
I_{\ket{\downarrow \cdots \downarrow},\text{Haar}}={2^{-L}}$. In contrast, in the localised phase, we expect that up to exponential corrections, the initial state spreads up to a finite distance $k$, so $\mathbb{U}^t\ket{\downarrow \cdots \downarrow}= \ket{\psi}\ket{\downarrow}^{\otimes(L-k)}$, which leads to an IPR constant in $L$. Namely $I_{\ket{\downarrow \cdots \downarrow},\text{loc}}\geq {2^{-k}}$ for all $L$. 

We computed $I_{\ket{\downarrow \cdots \downarrow}}$ numerically for several values of $a$, $\tau$, and $L$: our main numerical results are summarised in Fig.~\ref{fig:2dPD}. The behaviour of the IPR aligns remarkably well with the phenomenology of finite-time data. The bottom panel of the figure shows the IPR versus $L$ for different values $a$ and three choices of $\tau$. Identifying the localisation transition as the transition between constant and exponentially decaying IPR, we can estimate $a_c(\tau)$. The last two $\tau$ are similar in size, but they are respectively irrational and rational multiples of $2\pi$. We see that the difference between these two cases is stark also in this setting: the irrational $\tau/(2\pi)$ shows a transition at sizeable $a$, while rational $\tau/(2\pi)$ shows ergodic behaviour for the same choice of $a$. In the phase diagram, the rational $\tau/(2\pi)$ generate some irregular behaviour reminiscent of Arnold tongues~\cite{arnold1961small}. Some further discussion and additional finite-volume data is reported in the SM~\cite{Note1}. 

\textit{Discussion and Outlook.---} We introduced a discrete-time version of the Quantum East model~\cite{horssen2015dynamics} and analysed its localisation properties in real-time. Combining a perturbative analysis with exact numerics we identified a localisation transition taking place in this system despite the periodic drive: for couplings smaller than a critical value $a_c(\tau)$ the effect of a boundary perturbation remains localised in space, while it spreads ballistically for $a>a_c(\tau)$. This is also shown by a stark difference in the entanglement scaling (linear vs bounded), which is more marked than what is reported in other accounts of localisation in Floquet settings~\cite{ponte2015many, ray2018signature}. Interestingly, this transition has a non-analytic dependence on the Trotter time $\tau$ and takes place only when the latter is an irrational multiple of $2\pi$. In fact, irrational Trotter step and dynamical constraints can be identified as the two key mechanisms for the onset of localisation. 
To understand this one can imagine expanding the state of the system at time $t$ in the computational basis. Because of the dynamical constraints, there will be far fewer states appearing in this sum than those allowed by the locality of the interactions. Moreover, all configurations are dampened by a factor $(\tau a)^x$, where $x$ is the position of the up spin, that is further from the left boundary. This, however, is not enough to ensure that the configurations with large $x$ are suppressed --- i.e. localisation --- because each configuration can be reached in many different ways, i.e.,  by many different ``trajectories". This means that in the expansion of the state at time $t$ each configuration is multiplied by a ``combinatorial weight", which can in principle overcome the dampening. An irrational Trotter step avoids this by introducing destructive interference between the different trajectories, and hence, ensuring that the combinatorial weight never overcomes the exponential dampening. 
In the continuous time limit, $\tau\to0$, the dampening factor goes to 0 and the combinatorial weight diverges, therefore one has to combine the two effects. The outcome suggested by our analysis is that in this limit the system is localised for any $a<1$, in agreement with Ref.~\cite{pancotti2020quantum}. This is not in contrast with the statement that the Floquet Quantum East is localised only for irrational $\tau/(2\pi)$ as in this case the limits of $\tau\to 0$ and $t \to \infty $ do not commute~\cite{Note2}.

A natural question is what are the initial states for which localisation occurs. We note that our analysis can be repeated for all states written as tensor products of arbitrary finite-block states with an infinite block of down spins on the right. This is consistent with the local-operator spreading analogy discussed earlier, as such states are those corresponding to local operators. States not fitting this form evade our treatment, leaving their analysis for future research. Our expectations is that those states will not show localisation as in the continuous-time setting~\cite{pancotti2020quantum}. Indeed, they lack the first ingredient of the localisation mechanism we identified, i.e., the presence of dynamical constraints. 
A key future direction is the rigorous characterisation of the observed transition within the convenient discrete space-time setting introduced here.

\let\oldaddcontentsline\addcontentsline
\renewcommand{\addcontentsline}[3]{}

\textit{Acknowledgements.} P.K. thanks Giacomo Giudice for fruitful discussions.
B.B. was supported by the Royal Society through the University Research Fellowship No. 201101.
P.K. acknowledges financial support from the Alexander von Humboldt Foundation.
T.P. acknowledges the Program P1-0402 and Grants
N1-0219, N1-0233 of the Slovenian Research and Innovation Agency (ARIS).

\footnotetext[1]{See the Supplemental Material that contains: (i) A combinatorial calculation of $N(x,t)$. (ii) A perturbative analysis of brickwork partial norms $W(x,t)$. (iii) A perturbative analysis of $N(x,t)$ in the Trotter limit. (iv) A self-contained discussion of our TEBD algorithm. (v) Further TEBD data. (vi) Comparison of ladder and brick-wall settings. (vii) Detailed analysis of Fig.~4.  (viii) Further finite-volume data.}

\footnotetext[2]{To see this, we observe that to get to $\tau=0$ using rational multiples of $2\pi$ one has to take $\tau=2\pi/d$ with increasingly large $d\in \mathbb N$. In this case delocalisation is expected to emerge only for $t\gg d$ (cf.~\eqref{eq:qlucas}) leading to the claim.}
	
\bibliography{bibliography}

\let\addcontentsline\oldaddcontentsline

	\onecolumngrid
	\newpage

	\appendix
	\setcounter{equation}{0}
	\setcounter{section}{0}
	\setcounter{figure}{0}
	\renewcommand{\thetable}{S\arabic{table}}
	\renewcommand{\theequation}{S\thesection.\arabic{equation}}
	\renewcommand{\thefigure}{S\arabic{figure}}
	\setcounter{secnumdepth}{2}

\begin{center}
{\Large Supplementary Material for: \\ 
			``\titleinfo"	}
\end{center}
\tableofcontents

\section{Evaluation of $N'(x,t)$}

We begin introducing the amplitude corresponding to $N'(x,t)$ in Eq.~\eqref{eq:simp}, i.e. 
\be
\Ap(x,t) = \mel*{\underbrace{\uparrow\cdots \uparrow}_x \downarrow \cdots \downarrow }{\mathbb U_{\ell} (a,\tau)^t}{\downarrow \cdots \downarrow}.
\label{eq:amplitude}
\ee
Evaluating this amplitude at leading order is a simple combinatorial problem: we plug \eqref{eq:gateleading} and  \eqref{eq:interactionpicture} into \eqref{eq:amplitude} and count the number ways to flip $x$ spins in sequential order. For the $j$-th flip we get a factor $i a \tau e^{i \tau t_j}$, where ${t \leq t_j < t_{j-1}}$ is the time of the $j$-th flip. This gives 
 \be
\Ap(x,t) \simeq (-i)^x e^{i \tau t} (a \tau)^x K(x,t), 
\ee
where we introduced 
\be
K(x,t) \equiv  \sum_{t_1=1}^{t}\sum_{t_2=t_1+1}^{t} \cdots \!\!\!\!\!\sum_{t_x=t_{x-1}+1}^{t} e^{ i \tau (t_1+\cdots+ t_x)}.
\label{eq:Ksum}
\ee
It is simple to show that these objects fulfil the following recurrence relations 
\be
K(x,t) = K(x,t-1) + K(x-1,t-1) e^{i \tau t} , 
\ee
with $K(x> t,t) = 0$ and $K(0,0) =1$ which is solved by 
\be
K(x,t) = e^{-i \tau x (x+1)/2} \begin{pmatrix}
t \\
x
\end{pmatrix}_{\!\!\!q}\!\!. 
\ee 
Here we set $q\equiv e^{i \tau}$ and 
\be
\!\!\!\!\!\begin{pmatrix}
n \\
m
\end{pmatrix}_{\!\!\!q} \!\!=\!
\frac{[n]_q!}{[n-m]_q! [m]_q!},\,\, n\geq m,\quad
\begin{pmatrix}
n \\
m
\end{pmatrix}_{\!\!\!q} \!\!=0,\,\, n< m,
\ee
are $q$-deformed binomial coefficients defined in terms of $q$-deformed factorials
\be
[n]_q! = \prod_{k=1}^n [n]_q, \qquad [k]_q = \frac{1-q^k}{1-q}\,.
\ee
Putting all together we then find Eq.~\eqref{eq:WPT1st}.

\section{Perturbative analysis of $W(x,t)$}

The perturbative analysis of the main text can be directly performed also for the partial norms $W(x,t)$ in the brickwall formulation, Eq.~\eqref{eq:W}. Considering the interaction picture representation of the evolution operator \eqref{eq:Floquet} we have  
\be
\!\!\!\!{\mathbb U (a,\tau)^t} e^{i \tau t (1+\sum_{j} \! P_j)} \!\!=\!\!\prod_{k=0}^{t-1} \!\tilde{\mathbb U}_{\rm e}(\tau a, k\tau)\tilde{\mathbb U}_{\rm o}(\tau a, (k+1)\tau),
\label{eq:niceform}
\ee
where we introduced
\begin{align}
&\tilde{\mathbb U}_{\rm e}(a,\tau) = \tilde U(a, \tau)^{\otimes L},\\
&\tilde{\mathbb U}_{\rm o}(a,\tau)=  e^{- i a  X} \otimes \tilde U(a,\tau)^{\otimes (L-1)}\,,
\end{align}
and $\tilde U(a, \tau)$ is the one given in Eq.~\eqref{eq:Utilde}. Considering the amplitude 
\be
\Ap_W(x,t) = \mel*{\underbrace{\uparrow\cdots \uparrow}_x \downarrow \cdots \downarrow }{\mathbb U (a,\tau)^t}{\downarrow \cdots \downarrow} 
\ee
we find that it is again computed by counting the number ways to flip $x$ spins in sequential order. This time the $j$-th flip gives a factor 
\be
- i a \tau e^{-i \tau (t- t_j+{\rm mod}(j,2))}, 
\ee
where $t \leq t_j \leq t_{j-1}+{\rm mod}(j,2)$. Putting all together we find 
\be
\Ap_W(x,t) \simeq  (-i)^x e^{- i \tau [t(x+1)+\lceil x/2 \rceil ]}  (a \tau)^x C(x,t),
\ee
where we introduced 
\be
C(x,t)= \sum_{p_1=1}^{t}\sum_{p_2=p_1}^{t} \cdots \!\!\!\!\!\sum_{p_x=p_{x\!-\!1}+{\rm mod}(x,2)}^{t} \!\!\!\!\!\!  e^{ i \tau (p_1+\cdots+ p_x)}. 
\ee
The coefficients $C(x,t)$ fulfil the following recursive relation 
\begin{align}
& C(2x,t)= \,C(2x,t\!-\!1)+C(2x\!-\!1,t\!-\!1)e^{ i \tau t}+C(2x-2,t\!-\!1)e^{2 i \tau t},\\
& C(2x\!-\!1,t) =  C(2x\!-\!1,t\!-\!1)+C(2x\!-\!2,t\!-\!1)e^{i \tau t}\!\!\!,
\end{align}
with boundary conditions 
\begin{align}
&\!\!\!C(0,1) = 1, \quad C(1,1) =  e^{i \tau}, \quad C(2,1) =  e^{2 i \tau}\!\!,\\
&\!\!\!C(x<0 ,t) = 0, \quad C(x>2 t ,t) = 0. 
\end{align}
These equations are solved by 
\be
\!\!C(x,t) = q^{x/2+x^2/4+{\rm mod}(x,2)/4} 
\begin{pmatrix}
t+\lfloor x/2 \rfloor\\
x
\end{pmatrix}_{q=e^{i \tau}}\!\!\!.
\ee
Plugging back into \eqref{eq:simp} we finally find 
\be
W(x,t) \simeq (a \tau)^{2x} \left |C(x,t)\right |^2\,. 
\ee

\section{Perturbation theory in the Trotter limit}
Considering the Trotter limit of the time-evolution operator in the interaction picture we have 
\be
\lim_{\tau\to0}  {\mathbb U} (a,\tau)^{\frak t/\tau} = {\rm T}\!\exp\left[- i \int_0^{\frak t} {\rm d} s H_I(s,a)\right] e^{i \frak t \sum_{j} \! P_j },
\ee
where ${\rm T}\!\exp[\cdot]$ represents the time-ordered exponential and 
\be
H_I(s,a) = a \sum_j P_j X_{j+1} e^{- i s Z_{j+1}} + a(X_{1} e^{- i s Z_{1}} -\1),
\ee
is the coupling part of the Hamiltonian in the interaction representation. A standard expansion of the time-ordered exponential gives 
\begin{align}
\Ap(x,t) =& \mel*{\underbrace{\uparrow\cdots \uparrow}_x \downarrow \cdots \downarrow }{{\rm T}\!\exp\left[- i \int_0^{\frak t} {\rm d} s H_I(s,a)\right]}{\downarrow \cdots \downarrow}\notag\\
=& (-i)^x a^x \mel*{\underbrace{\uparrow\cdots \uparrow}_x \downarrow \cdots \downarrow }{ X_x P_{x-1}\ldots X_2 P_1 X_1}{\downarrow \cdots \downarrow} \int_0^{\frak t} {\rm d}t_x \int_0^{t_x} {\rm d}t_{x-1}\cdots \int_{0}^{t_2} {\rm d}t_1 e^{i (t_x+\ldots +t_1)}+O(a^{x+1})\notag\\
=& (-2i a)^x \frac{\sin(\frak t/2)^x}{x!}\,.
\end{align}
Plugging back into Eq.~\eqref{eq:simp} we find Eq.~\eqref{eq:tauto0}.

\section{TEBD algorithm for conditional ladder circuit}

To simulate the dynamics in the Floquet Quantum East circuit in the ladder formulation we use the standard time-evolved block decimation (TEBD) algorithm~\cite{vidal}, which is ideally suited for this particular application. We write the state
\be
\ket{\psi(t)} = \mathbb U_\ell^t \ket{\downarrow\downarrow\cdots}
\ee
which we represent as a matrix product state
\be
\ket{\psi(t)} = \sum_{s_1,\ldots,s_t\in\{\uparrow,\downarrow\}} A^{(1,t)}_{s_1} \Lambda^{(1,t)}  A^{(2,t)}_{s_2} \Lambda^{(2,t)} \cdots A^{(t,t)}_{s_t} \ket{s_1 s_2\cdots s_t
\downarrow\downarrow\cdots}.
\ee
$A^{(x,t)}$ are $d^{(x-1,t)} \times d^{(x,t)}$ dimensional matrices, which at any position $x\in\{1,\ldots,t\}$ and instant of time $t\in\mathbb N$ satisfy the (right) orthogonality 
relations
\be
\sum_{s\in\{\uparrow,\downarrow\}} A^{(x,t)}_s \bigl[\Lambda^{(x,t)}\bigr]^2 \bigl[A^{(x,t)}_s\bigr]^\dagger = I,
\ee
where $\Lambda^{(x,t)}$ are $d^{(x,t)}$ 
dimensional diagonal matrices containing (nonzero) Schmidt coefficients 
\be
\Lambda^{(x,t)}={\rm diag}\{\sigma^{(x,t)}_n|n=1,2,\ldots d^{(x,t)}\}\,.
\ee
Namely, they are square roots of the elements of entanglement spectrum $[\sigma^{(x,t)}_n]^2$ for the bipartition  $[1,2,\ldots t] \cup [t+1,t+2\ldots]$.\\

\noindent For convenience and consistency we take $d^{(0,t)}:=1$, $d^{(t,t)}:=1$, $\Lambda^{(0,t)}:=1$, $\Lambda^{(t,t)}:=1$. Note that due to the normalization of the state $\ket{\psi(t)}$ we have $\tr [\Lambda^{(x,t)}]^2 = 1$.\\

\noindent In the $t$-th time step of the algorithm, having already computed the tensors $A^{(x,t-1)}_s,\Lambda^{(x,t-1)}$, 
$x=1,\ldots t-1$, we start applying the gates of the ladder propagator $\mathbb U_\ell$ from the right. We first trivially expand the support of the state $\ket{\psi(t-1)}$
one site to the right, by placing an explicit down-spin at place $x=t$ introducing an $1\times 1$ impurity matrix  $B^{(t)}_s = \delta_{s,0}$, i.e.
\be
\ket{\psi(t-1)} = \sum_{s_1,\ldots,s_t\in\{\uparrow,\downarrow\}} A^{(1,t-1)}_{s_1} \Lambda^{(1,t-1)}  A^{(2,t-1)}_{s_2} \Lambda^{(2,t-1)} \cdots \Lambda^{(t-1,t-1)}B^{(t)}_{s_t} \ket{s_1 s_2\cdots s_t
\downarrow\downarrow\cdots}.
\ee

\tikzstyle{RectObject}=[rectangle,fill=white,draw,line width=0.5mm]

\begin{figure}
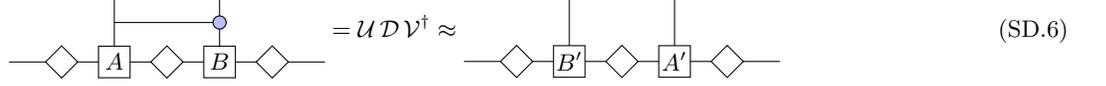

    \centering
    \begin{align}
     \btp
\XeBig{2}{0.75}
\draw (-1,0)-- (5,0);
\node[diamond,draw, fill=white] (d1) at (0,0) {};
\node[diamond,draw, fill=white] (d2) at (2,0) {};
\node[diamond,draw, fill=white] (d3) at (4,0) {};
\draw[draw=black, fill=white]  (-0.3+1,-0.3) rectangle (.3+1,.3);
\draw[draw=black, fill=white]  (-0.3+3,-0.3) rectangle (.3+3,.3);
\node (A) at (1,0) {$A$};
\node (B) at (3,0) {$B$};
%
%
    \etp
    = \mathcal U\, \mathcal D\, \mathcal V^{\dagger} 
    \approx	
\btp
\draw (-1,0) --(5,0);
\node[diamond,draw, fill=white] (d1) at (0,0) {};
\node[diamond,draw, fill=white] (d2) at (2,0) {};
\node[diamond,draw, fill=white] (d3) at (4,0) {};
\draw (1,0)--(1,1.25);
\draw (3,0)--(3,1.25);
\draw[draw=black, fill=white]  (-0.3+1,-0.3) rectangle (.3+1,.3);
\draw[draw=black, fill=white]  (-0.3+3,-0.3) rectangle (.3+3,.3);
\node (A) at (1,0) {$B'$};
\node (B) at (3,0) {$A'$};
%
%
    \etp
    \end{align}
    \caption{Here we show the elementary iterative step of our TEBD algorithm. Diamonds denote the $\Lambda$ matrices of Schmidt coefficients.
    First we construct the matrix $M$ with
    tensors
    $A=A^{(x,t-1)}$ and $B=B^{(x+1)}$, which we then write in the canonical singular value decomposition. Then we truncate the singular values, put them in the new $\Lambda^{(x,t)}$, and define new tensors $A'=A^{(x+1,t)}$ and $B'=B^{(x)}$.
    In this way we applied one local conditional gate and moved the impurity tensor $B$ one place left.
    }
    \label{fig:TEDB}
\end{figure}

\noindent Then, for $x=t-1,t-2,\ldots 1$, we do the following (see also the diagram in Fig.~\ref{fig:TEDB})
\begin{itemize}
\item[1.] We form a $2 d^{(x-1,t-1)} \times 2 d^{(x+1,t)}$ matrix by multiplying local matrices and
performing a local conditional gate
\be
M_{(s,n),(s',n')} = \sum_{s''}
(\delta_{s,\downarrow}\delta_{s',s''}+
\delta_{s,\uparrow} u_{s',s''})
\big[\Lambda^{(x-1,t-1)} A_s^{(x,t-1)} \Lambda^{(x,t-1)} B^{(x+1)}_{s''} \Lambda^{(x+1,t)}\bigr]_{n,n'}\,,
\ee
where $u$ is the single-qubit part of of full 2-qubit conditional gate $U$ (\ref{eq:localgate})
\be 
u =\exp(-i \tau( a X - I)).
\ee
\item[2.] We compute the canonical singular value decomposition of $M$
\be
M =: \mathcal U\, \mathcal D\, \mathcal V^{\dagger}
\ee
where we keep only $d^{(x,t)}$ singular values (elements of the diagonal matrix $\mathcal D$) larger than some prescribed truncation accuracy $\varepsilon$. Thereby we assigning 
\bea
&& \bigl[B^{(x)}_s\bigr]_{n,n'} := \bigl[\Lambda^{(x-1,t-1)}_{n,n}\bigr]^{-1} \mathcal U_{(s,n),n'},\\
&& \Lambda^{(x,t)} := {\rm diag}\{\mathcal D_{n,n}|n=1,\ldots,d^{(x,t)}\},\\
&& 
\bigl[A^{(x+1,t)}_s\bigr]_{n,n'} := \bigl[\Lambda^{(x+1,t)}_{n,n}\bigr]^{-1} \mathcal V^*_{(s,n),n'}.
\eea
\end{itemize}
This means that in each iteration we move the defect matrices $B^{(x)}_s$ one step to the left, maintaining the canonical Schmidt orthogonal form of the matrix product state. At the end of the loop, we set 
\be
A^{(1,t)}_s := \sum_{s'} u_{s,s'} B^{(1)}_{s'},
\ee
which applies the local unconditional gate $u$ at the left end of the circuit (see Fig.~\ref{fig:circuit}).
\\\\
We simulated dynamics of Floquet Quantum East chain using TEBD algorithm in the localised regime with negligible truncation error, setting $\varepsilon$ between $10^{-6}$ and $10^{-10}$. This meant in practice that dynamical bond dimensions $d^{(x,t)}$ never grew to more than a few hundred for the data shown in this paper. On the other hand, in the ergodic regime, $d^{(x,t)}$ quickly grew so that TEBD could only reach times comparable to those accessible to exact simulation $t\approx 30-40$.

\section{Additional data from TEBD (and exact) simulations of infinite systems at finite time}

\subsection{Quantitative check of perturbative analysis}

\begin{figure}     \hspace{-5mm}\includegraphics[width=\textwidth]{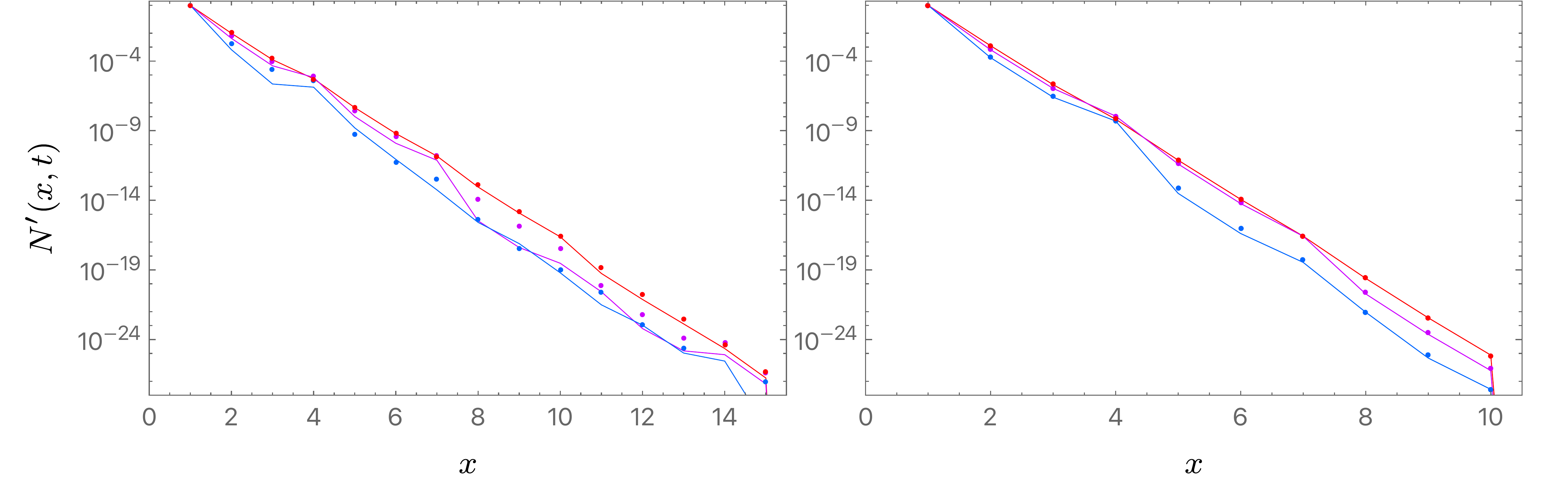}
    \caption{
    Quantitative match with perturbation theory:
    Domain wall profile components $N'(x,t)$ for small $a$, 
    $a=0.06$ (left), $a=0.02$ (right), and $\tau=(\sqrt{5}-1)\pi/2$, and three different $t=25$ (red), $t=35$ (magenta), $t=45$ (blue). Bullets denote leading order perturbative formula (\ref{eq:WPT1st}).
 }
 \label{fig:pert}
\end{figure}

For very small coupling parameter $a$ the leading order perturbative prediction (\ref{eq:WPT1st}) for the domain wall components $N'(x,t)$, gives even quantitatively correct result. We show in Fig.~\ref{fig:pert} the comparison between perturbative prediction and TEBD data for $a=0.02$ and $a=0.06$
and irrational Trotter time $\tau=(\sqrt{5}-1)\pi/2$. We find indeed that for times $t \lesssim 1/a$
the agreement is even quantitative, whereas for longer times $t \gg 1/a$ the agreement is still qualitative, namely the overall decay of $N'(x,t)$ seems to be correctly captured by perturbation theory for all times $t$.

\subsection{Critical coupling parameter $a_c(\tau)$ for smaller $\tau$}

\begin{figure}     \hspace{-5mm}\includegraphics[width=0.95\textwidth]{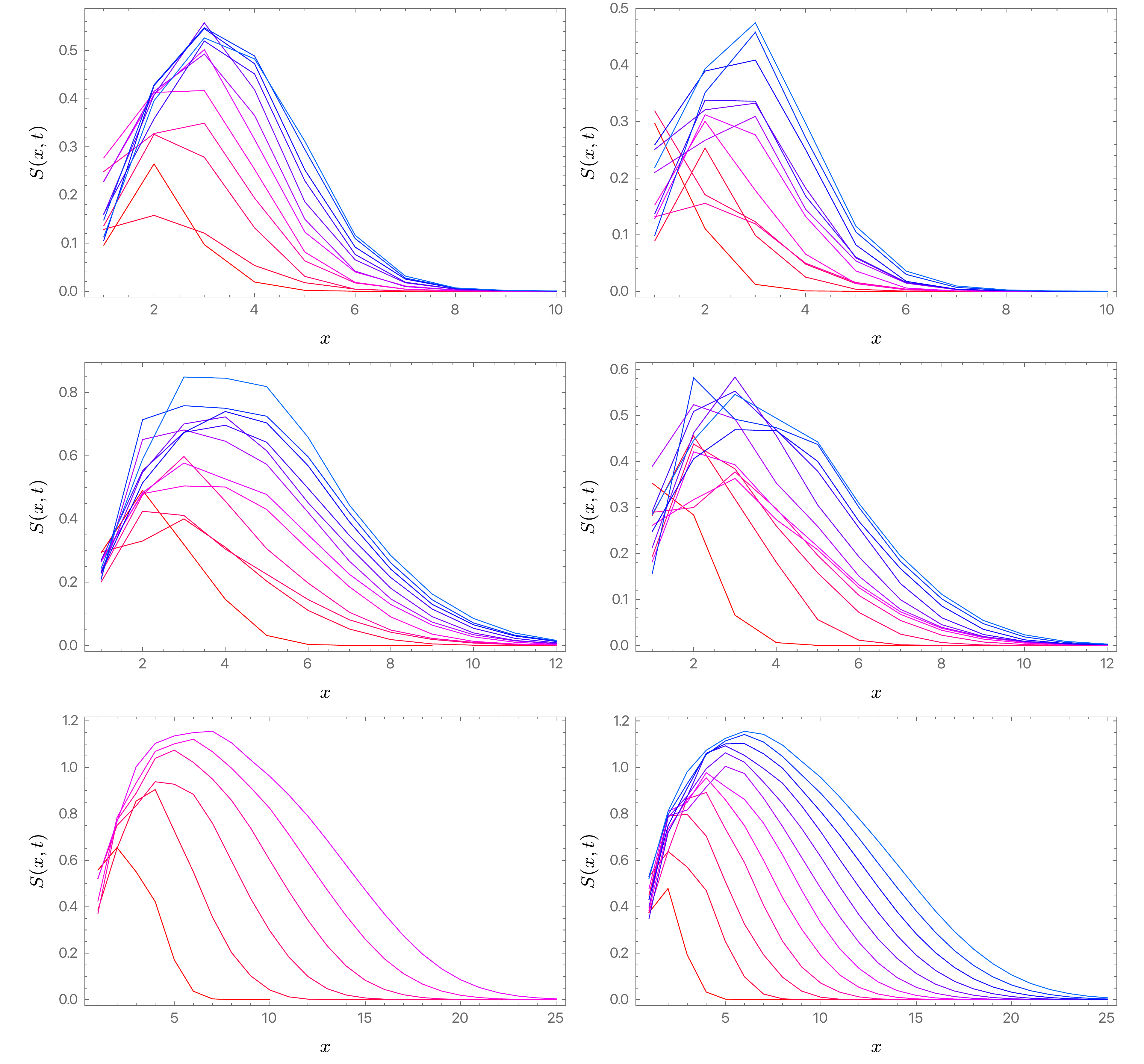}
    \caption{
    Entanglement entropy dynamics $S(x,t)$,
    for $t=20,30,\ldots,120$ (red to blue curves) approaching the localisation transition for smaller values of $\tau$,
    specifically $\tau=(\sqrt{5}-1)\pi/8$ (left column panels) and $\tau=(\sqrt{5}-1)\pi/16$ (right column panels), and three different values of $a$, $a=0.6$ (top row panels), $a=0.8$ (middle row panels), $a=1.0$ (bottom row panels). Note that only in the bottom row panels we see clearly the linear growth of entanglement entropies, signalling ergodic dynamics, so the transition should appear for $0.8 < a_c < 1.0$ for both values of $\tau$ (compare against the phase diagram in the main text).    
 }
 \label{fig:other}
\end{figure}

In order to verify our perturbative prediction (\ref{eq:actworegimes})
for the critical coupling constant $a_c(\tau)$ we also investigate the dynamics of entanglement entropies $S(x,t)$ as function of $a$ for values of $\tau$ considerably smaller than those shown in Fig.~\ref{fig:infTloc}. Specifically, in Fig.~\ref{fig:other} we study cases of $a=0.6,0.8,0.10$ and $\tau=(\sqrt{5}-1)\pi/8$ and $\tau=(\sqrt{5}-1)\pi/16$ clearly suggesting the transition to lie in the interval $a_c\in[0.8,1]$ for both small values of $\tau$, in qualitative agreement with the prediction (\ref{eq:actworegimes}) and even quantitatively agreing with the IPR phase diagram (\ref{fig:2dPD}).

\subsection{Schmidt (entanglement) spectra}

\begin{figure}     \hspace{-5mm}\includegraphics[width=\textwidth]{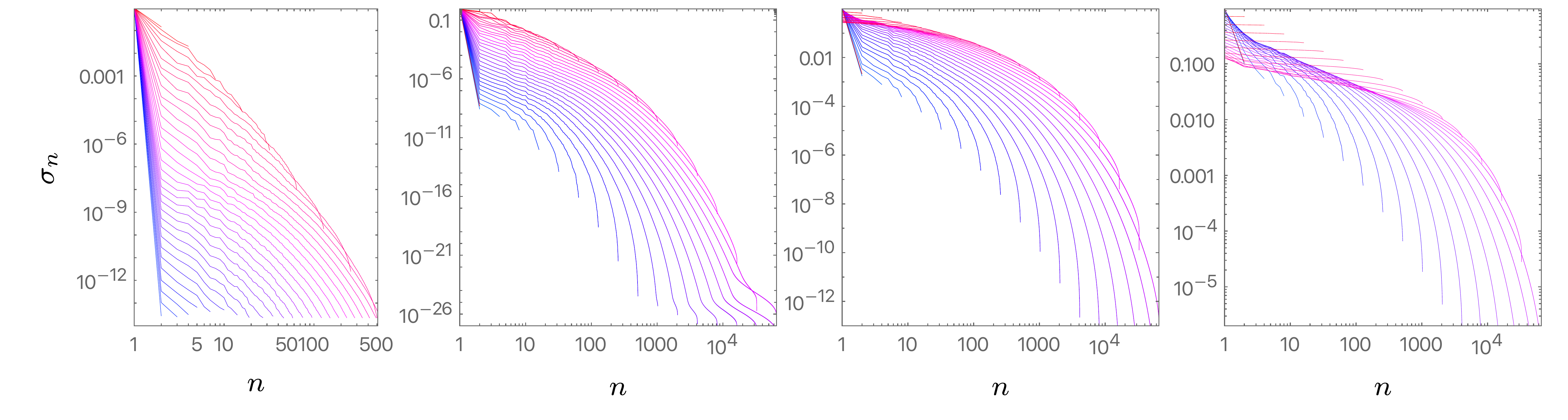}
    \caption{
    Entanglement (Schmidt) spectra:
    We show Schmidt spectra for $\tau=(\sqrt{5}-1)\pi/2$, $t=36$, and different $a=0.2,0.3,0.5,1.0$ (left to right panels).
    Red to blue curves correspond to cuts from $x=2,3,\ldots,35$. Schmidt coefficients are approximately log normal distributed in the ergodic region.}
 \label{fig:schmidt}
\end{figure}

It is also instructive to check the scaling of the Schmidt (or entanglement) spectra $\{\sigma^{(x,t)}_n\}$
across the localisation transition for cuts at different positions $x$ and sufficiently long time $t$. This is shown in Fig.~\ref{fig:schmidt} for $a=0.2$ (localised), and $a=0.3, 0.5, 1.0$ (ergodic regime), and fixed $\tau=(\sqrt{5}-1)\pi/2$, $t=36$.  We see that in the localised regime and for large enough $x$ the Schmidt values decay very rapidly in $n$. This signals lack of entanglement between a small region around the boundary and the rest of the system. As expected, this behaviour is instead not observed in the ergodic regime. Note that the data for ergodic regime are computed by exact simulation (not TEBD) of Hilbert space vectors in $2^{36}$ dimensional Hilbert space.

\subsection{(Spatio-)Temporal behaviour of entanglement entropies and partial norms}

\begin{figure}     
\begin{center}
\includegraphics[width=0.9\textwidth]{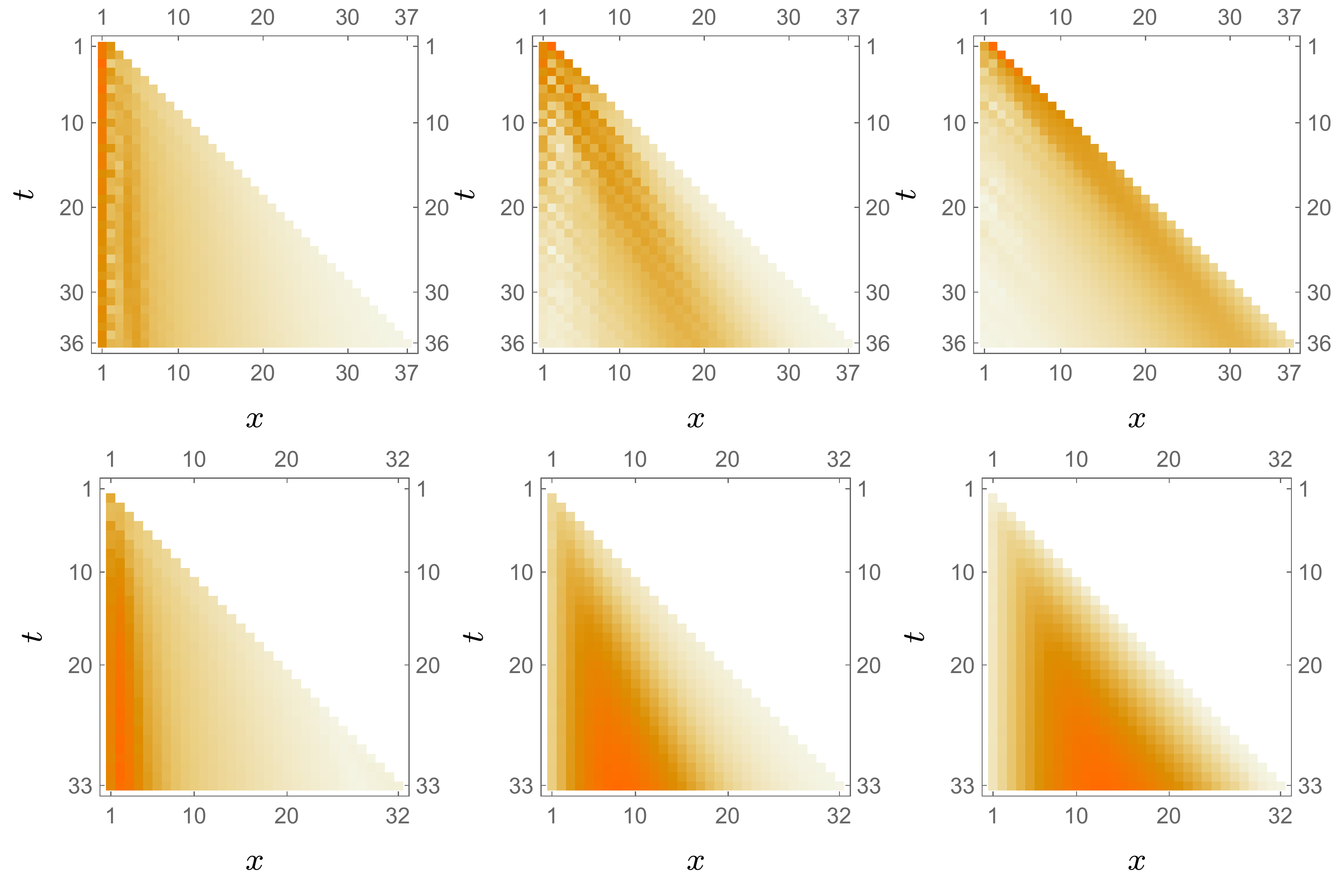}

\includegraphics[width=0.4\textwidth]{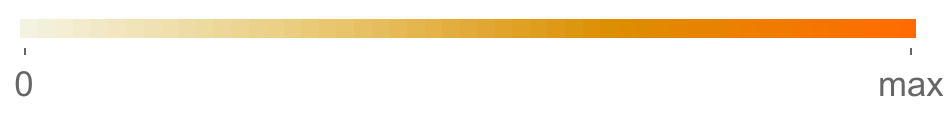}
\end{center}
    \caption{ Spatio-temportal dynamics of partial norms $N(x,t)$ --- top panels, and von Neumann entanglement entropy $S(x,t)$ --- bottom panels, for $\tau=(\sqrt{5}-1)\pi/2$, and $a=0.3$ --- left, $a=0.5$ --- middle, and $a=1.0$ --- right panels, all in the ergodic regime.  
    The horizontal bar indicates the colour code uniformly spanning, within each panel, from minimal value $0$ to maximal values, which read, in respective panels,
    $0.94$, $0.68$,
    $0.87$ (top),
    $1.03$, $4.30$,
    $7.49$ (bottom).    
    Data are computed by exact simulation on $36$ qubits.
 }
 \label{fig:heatergodic}
\end{figure}

\begin{figure}
\begin{center}
\includegraphics[width=0.95\textwidth]{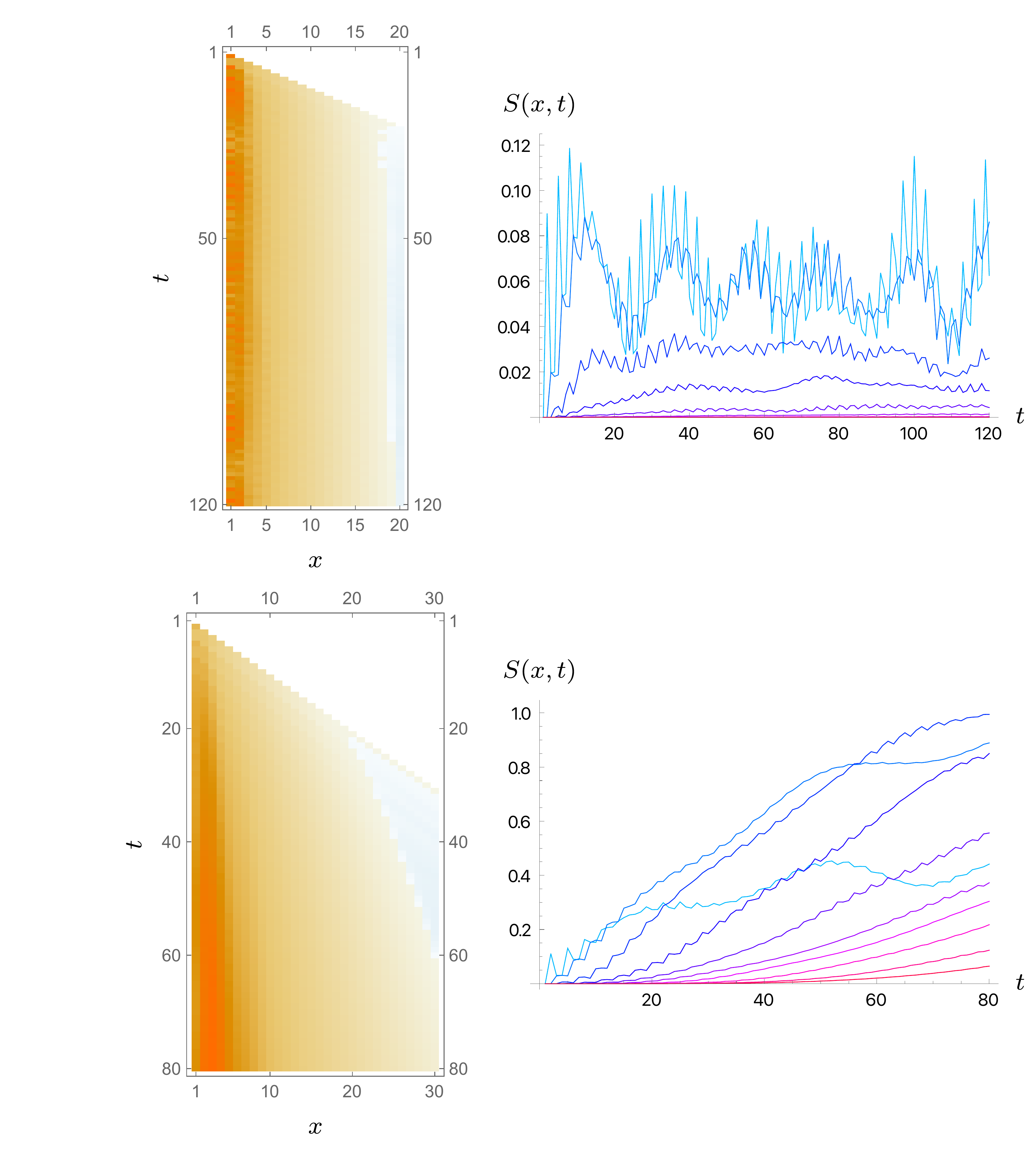}

\includegraphics[width=0.4\textwidth]{bar.pdf}
\end{center}
    \caption{
    (Spatio-)temporal dynamics of von Neumann entropy $S(x,t)$, heatmaps --- left panels, and time-dependence for fixed cuts at $x=1,2,\ldots$ (blue to red curve) --- right panels.
    Top panels show localised regime for $a=0.2$, $\tau=(\sqrt{5}-1)\pi/2$, while bottom panels show resonant (non-localised) regime $a=0.2$,
    $\tau=2\pi/3$. Data is computed by TEBD with maximal truncation error $\varepsilon=10^{-10}$.
    Again, horizontal colour bar designates linear colour code spanning values from $0$ to $0.12$ (top) $1.0$ (bottom).}
 \label{fig:heatlocalized}
\end{figure}

In order to corroborate the information displayed in Figs. 2 and 3 of the main text, here we provide additional data on the behaviour of the partial norms $N(x,t)$ and the von Neumann entanglement entropies $S(x,t)$. Specifically, we show the evolution of these quantities for fixed $x$ and their spatiotemporal heat-map plots. 
In Fig.~\ref{fig:heatergodic} we display $N(x,t)$, $S(x,t)$ in the ergodic regime $a=0.3,0.5,1.0$, $\tau=\sqrt{5}-1)\pi/2$, clearly displaying the lightcone effect.
In Fig.~\ref{fig:heatlocalized} we report the entropy heatmap $S(x,t)$ in the localised regime
$a=0.2$, $\tau=(\sqrt{5}-1)\pi/2$ and in the corresponding resonant case $\tau=2\pi/3$. In the former case we see that the entanglement entropy oscillates around a constant value, while in the latter we see a slow but persistent entanglement growth.

\clearpage
\section{Ladder vs brick-wall circuit}
\begin{figure}[t]
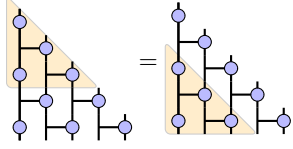

    \centering
    \begin{align}
    \btp
        \draw[ fill=myorange, rounded corners=2, opacity=0.2] (0,2) -- (2-1/4,1/4) -- (0,1/4) --cycle;
    \draw[thick] (1/4,-3/4) -- (1/4,7/4);
    \foreach \x in {3,1}{
    \Xe{\x/2}{0}
    \Xe{\x/2+1/2}{-1/2}
    }
    \Xe{2/2}{1/2}
    \Xe{1/2}{2/2}
    \XeOne{0}{-1/2}
    \XeOne{0}{1/2}
    \XeOne{0}{3/2}
    \etp
    =
    \btp
    \draw[ fill=myorange, rounded corners=2, opacity=0.2] (-1/2,2-2) -- (3/2-1/4,1/4-2) -- (-1/2,1/4-2) --cycle;
        \draw[thick] (1/4-1/2,-7/4) -- (1/4-1/2,3/4);
    \foreach \x in {3,...,0}{
    \Xe{\x/2}{-\x/2+0/2}
    }
    \XeOne{-1/2}{1/2}
    \Xe{1/2}{-3/2}
    \Xe{-0/2}{-2/2}
    \XeOne{-1/2}{-1/2}
    \XeOne{-1/2}{-3/2}
    \etp\notag
    \end{align}
    \caption{
    Similarity transformation $\mathbb{S}$ (shaded region) between the $\mathbb{U}$ and $\mathbb{U}_\ell$: $\mathbb{S}\mathbb{U}=\mathbb{U}_{\ell} \mathbb{S}$. 
    }
    \label{fig:similarity}
\end{figure}

In Fig.~\ref{fig:similarity} we show explicitly the similarity transformation, as a piece of quantum circuit, between the brick-wall and the ladder propagators of the Floquet Qantum East model.

\section{Detailed discussion of the Fig.~4 from the main text}

\begin{figure}[t]
    \centering
    \includegraphics[width=0.47\textwidth]{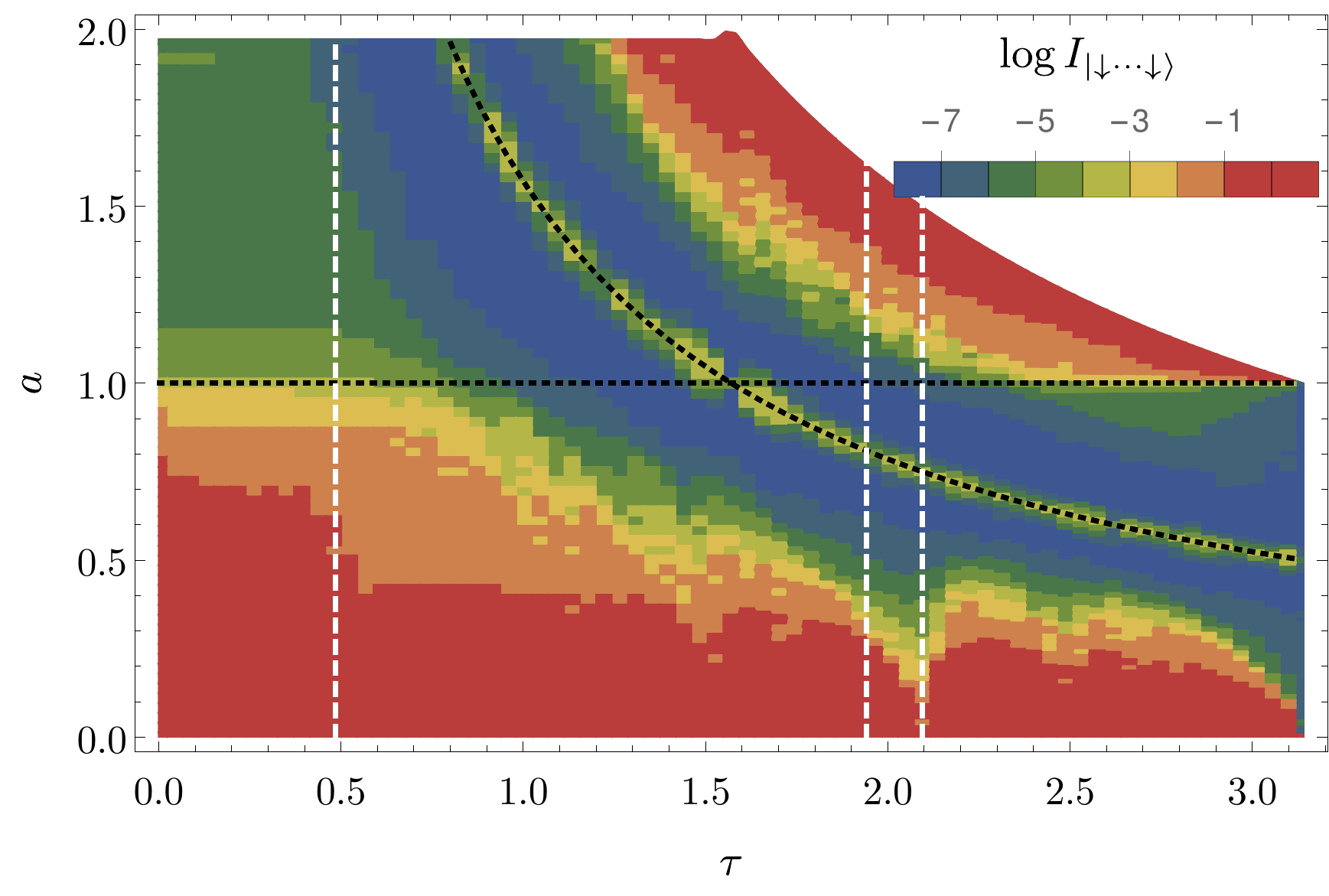} 
    \includegraphics[width=0.485\textwidth]{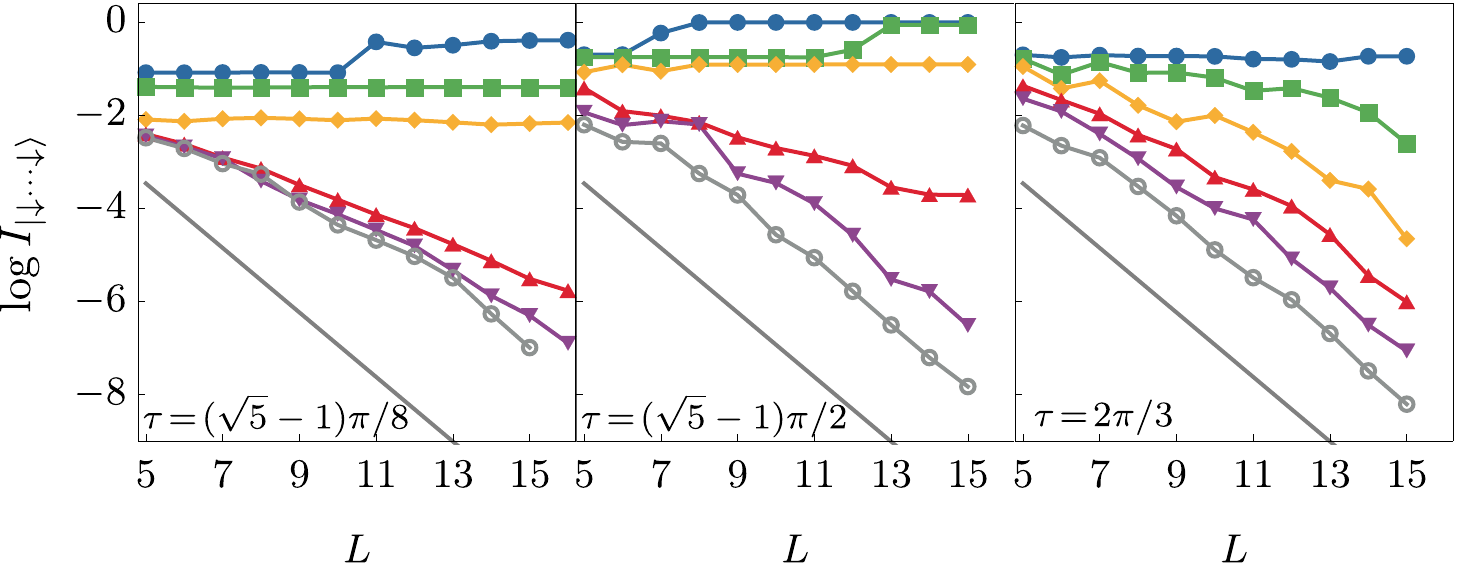}
    \caption{
    (Left) Logarithm of the IPR at $L=13$ as a function of $\tau$ and $a$. We see that for small $\tau$ the transition is at $a=1$, whereas for bigger $\tau$ the transition is at $a\propto 1/\tau$. With the white dashed lines we show the three values of $\tau$ considered in the bottom panel. 
    (Right) Logarithm of the IPR versus $L$ for three values of $\tau$ and several values of $a$ ($0.5,0.7,\dots,1.5$ top to bottom for the first plot and $0.05,0.15,\dots,0.55$ for the second and third). The grey solid line corresponds to fastest decay $2^{-L}$ for random eigenstates. For the first two values of $\tau$, we can estimate the transition between constant and exponential decay at $a=1$ and $a=0.3$. The third $\tau$ is a rational multiple of $2\pi$, and its transition occurs at a much smaller $a$.}
    \label{fig:2dPDSM}
\end{figure}

Here we explain some finer features of Fig. 4 from the main text, which we show again as Fig.~\ref{fig:2dPDSM}. 
First of all, we observe that there is not only a localised region around $a=0$, but also around $a =\pi/\tau $. This follows from the periodicity of the parametrisation of the gate:
\begin{align}
    U(a=\pi/\tau+\varepsilon,\tau)=e^{- i ((\pi+/\tau \varepsilon) P\otimes X- \tau P\otimes I)}=e^{- i (\tau \varepsilon P\otimes X- (\tau+\pi) P\otimes I)}\,.
\end{align}
The model is localised for small $\varepsilon$ for the same reason as it is localised for small $a$.

Secondly, the IPR is enhanced at $a=\pi/(2\tau)$, which is denoted by the black dotted line in the figure. There the local gate is
\begin{align}
    U(a=\pi/(2\tau),\tau)=  P\otimes X e^{ i (\tau-\pi/2)} + (\1-P) \otimes \1\,.
\end{align}
This is CNOT gate with an additional control phase. When the global evolution implemented by this gate repeatedly acts on our specific initial state $\ket{\downarrow\dots\downarrow}$, it always produces a state in the computational basis times a phase, and never a superposition. 
This is clearly non-ergodic dynamics, which results in enhanced IPR. Moreover, it has a ballistically moving front, so it is not localised.


Thirdly, let us discuss the signatures of the rational and irrational
nature of $\tau$ in the figure. As the figure is done with finite system size $L=13$, there is a finite resolution of $\tau$. The behaviour is very similar to classical examples, such as Arnold tongues for the standard circle map. 
Therefore, the figure does not look very different if we sample $\tau$ by irrational or rational steps of similar size, e.g. $\Delta \tau = 0.0436$ vs $\Delta \tau = \pi/72$ (here we adopt the latter). The values of $\tau$ that show the most irrational behaviour are fractions of the golden angle, since their finite fractional approximation of fixed order are the least accurate.
In the figure, we indicate two such values with white line $\tau={(\sqrt{5}-1)\pi}/{8}$ and $\tau={(\sqrt{5}-1)\pi}/{2}$.
The signature of rational behaviour is most pronounced around a small-denominator fraction of $\pi$ such as $\tau={2\pi}/{3}$ indicated by the third white line in the figure. Due to the finite size effects, the system looks localised for $a$ small enough even in this case, but the transition is shifted to much smaller values of $a$.

In the panel on the right, we see that IPR jumps from one constant value to bigger one by increasing the system size. It suggests that the system gets more localised upon increasing the system size, contrary to what happens in the usual examples of many-body localisation.
Nevertheless, this is a finite size effect which we do not understand well. Let us stress that for the indication of localisation, it is important how IPR changes with the system size, not its value per se.

\section{Additional data for IPR from exact diagonalization of finite systems}\label{app:IPR}

\begin{figure*}
    \centering
    \includegraphics[width=0.45\textwidth]{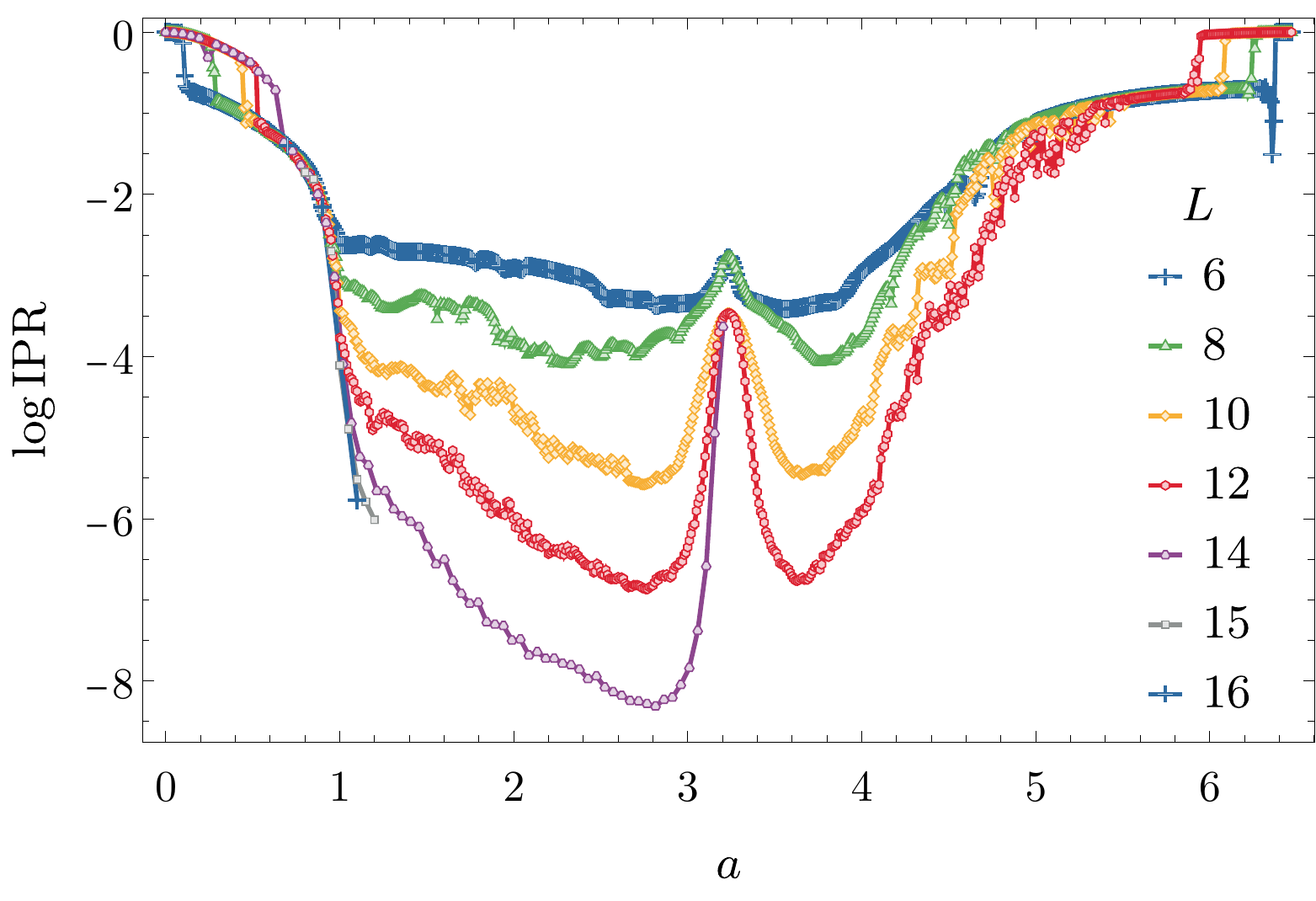}
    \includegraphics[width=0.45\textwidth]{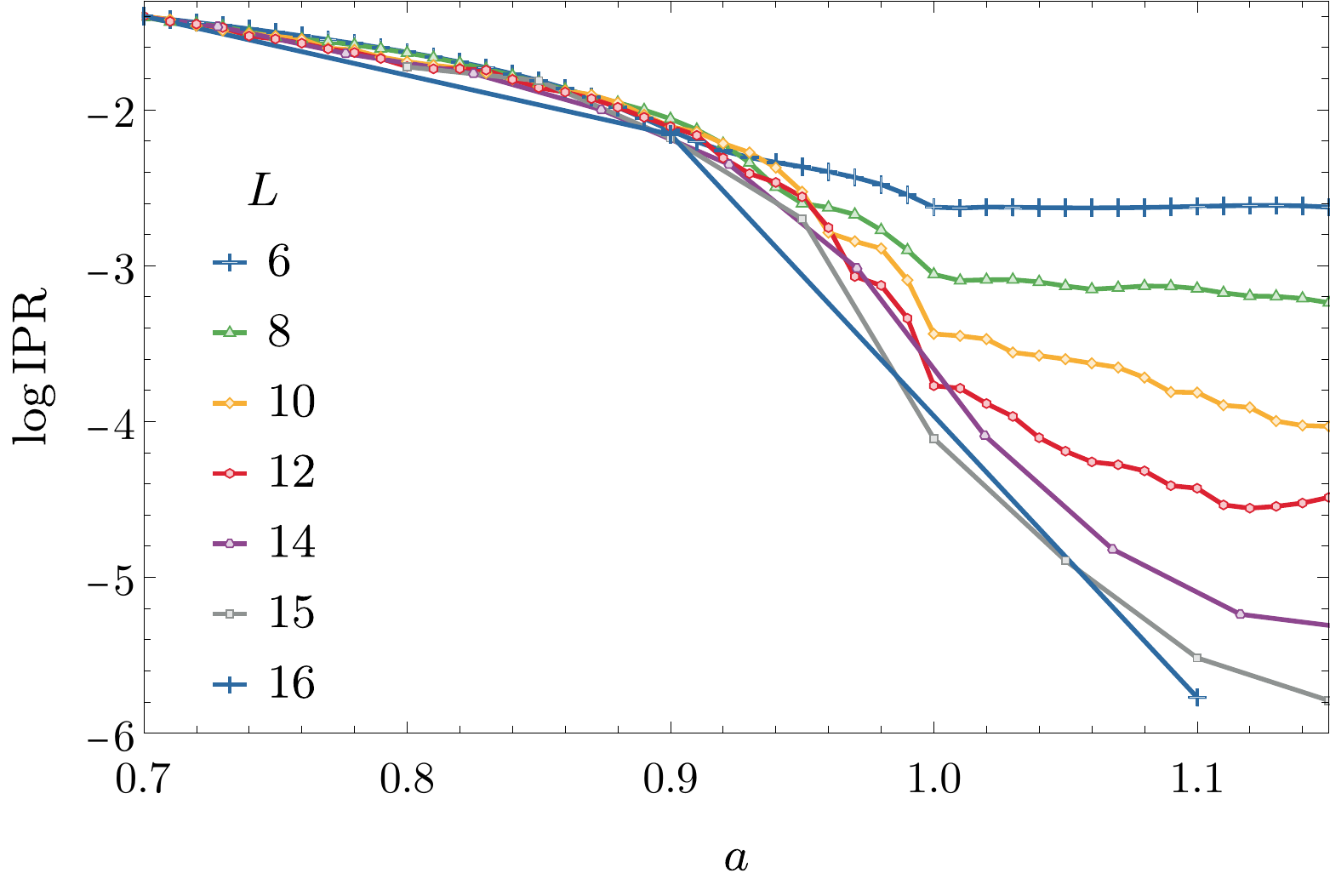}
    \includegraphics[width=0.45\textwidth]{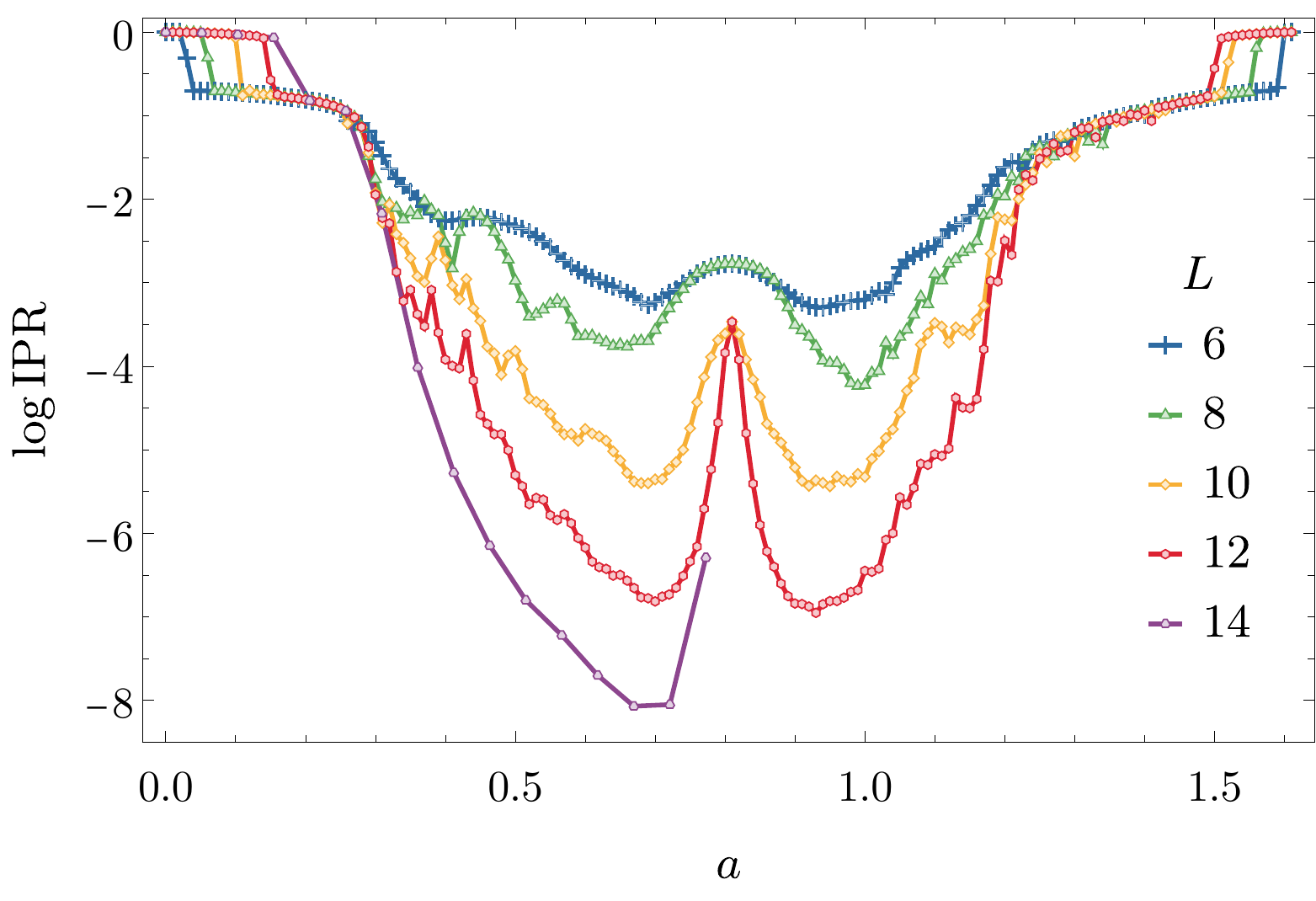}
    \includegraphics[width=0.45\textwidth]{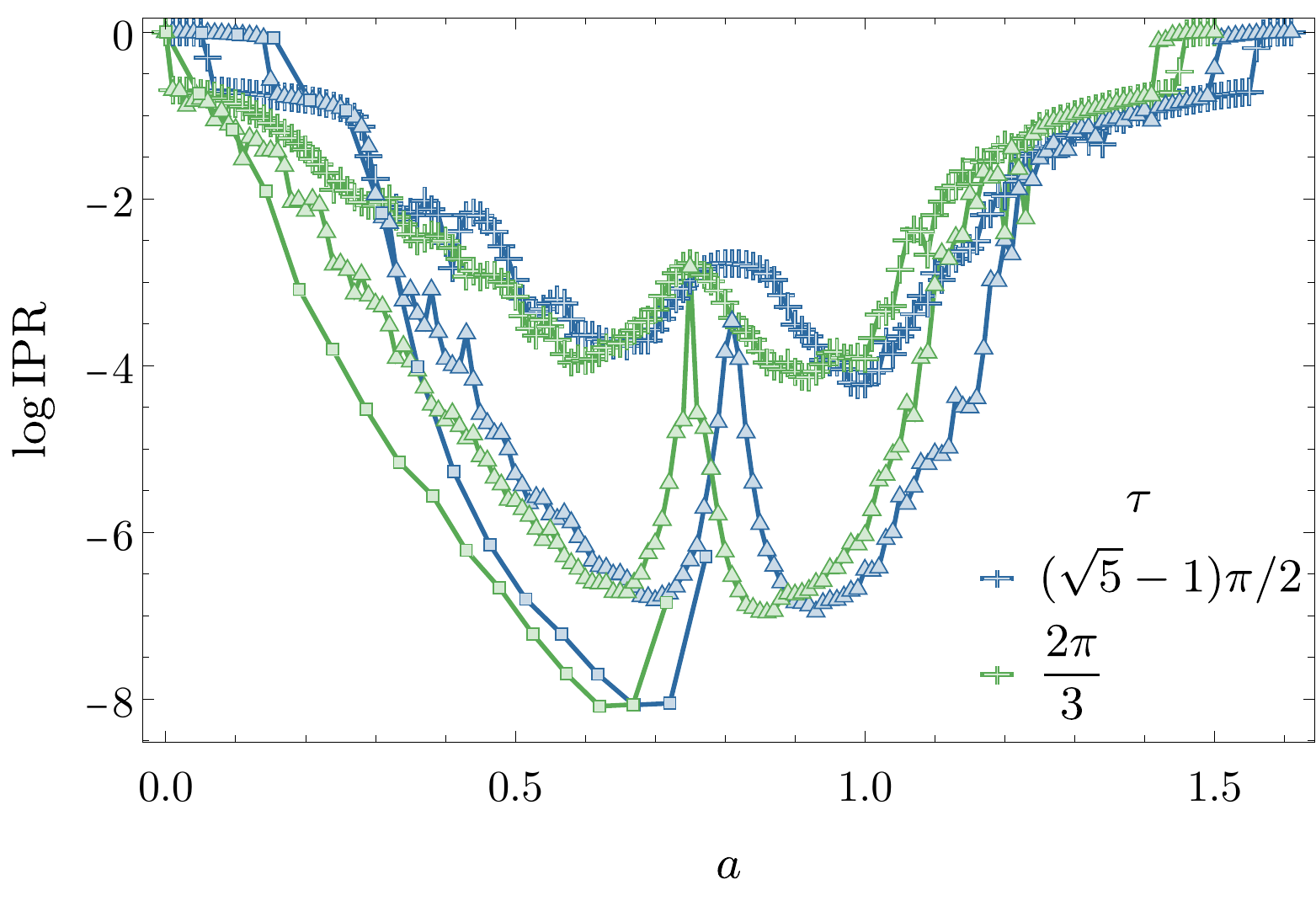}
    \caption{
    (Top) Logarithm of IPR versus $a$ at $\tau=(\sqrt{5}-1) \pi/8$ for different system sizes $L$.  We see the  the transition close to $a=1$. On the right we zoom on the collapse of data for different system sizes close to transition. (Bottom left)  Same as above with $\tau=(\sqrt{5}-1) \pi/2$. (Bottom right) A comparison between irrational $\tau=(\sqrt{5}-1)\pi/2 \sim 1.94161$ (blue) and rational $\tau=\frac{2\pi}{3}\sim 2.0944$ (green) for system sizes $L=8, 12, 14$ (cross, triangle, square). Even thought these $\tau$ are close, the transition of the irrational one is close to $a=0.3$ versus the rational one shows IPR shrinking with system size for much smaller $a$.
    }
    \label{fig:vsa}
\end{figure*}

In the main text we showed how IPR scales with $L$, and with $\tau$ and $a$. It is also illustrative to look at the scaling with $a$ for a fixed $\tau$ and a few different values of $L$, which we report in Fig.~\ref{fig:vsa}.
We see that there is an almost $L$ independent part where the IPR decreases significantly with increasing $a$. This could be interpreted as the region where the localisation length $\xi$ increases with increasing $a$, resulting in decreasing IPR $\sim 2^{-\xi}$. When $\xi$ increases beyond $L$, we see $L$ dependent values of IPR.

\end{document}